\documentclass{elsart}
\usepackage{graphicx,amssymb,amsmath}

\newcommand{\la}{\langle}
\newcommand{\ra}{\rangle}
\newcommand{\nn}{\nonumber}
\newcommand{\mR}{\mathcal{R}}
\newcommand{\mQ}{\mathcal{Q}}
\newcommand{\mG}{\mathcal{G}}

\newcommand{\om}{\mathop{\hat{\Omega}}}

\begin{document}
\begin{frontmatter}
\title{A proof of the reggeized form of amplitudes with quark
exchanges\thanksref{RFBR}}
\thanks[RFBR]{Work supported by the Russian Fund of Basic Researches,
projects 03-02-16529-a, 04-02-16685-a.}
\author{A.V.~Bogdan}
\ead{A.V.Bogdan@inp.nsk.su}
 and \author{V.S.~Fadin}\ead{fadin@inp.nsk.su}
\address{Budker Institute of Nuclear Physics, 630090 Novosibirsk, Russia}

\begin{abstract}
A complete proof of the quark Reggeization hypothesis in the
leading logarithmic approximation  for any quark--gluon inelastic
process in the multi--Regge kinematics  in all orders of $\alpha_s$
is given. First, we show that the multi--Regge form of QCD
amplitudes is guarantied if a set of conditions on the Reggeon
vertices and the trajectories is fulfilled. Then, we examine these
conditions and show that they are satisfied.
\end{abstract}
\end{frontmatter}

\section{Introduction}

Along with the Pomeron, which appears in QCD as a
compound state of two Reggeized gluons \cite{BFKL}, the
hadron phenomenology requires Reggeons, which can be
constructed as  colorless states of Reggeized quarks and
antiquarks. It demands further development of the theory
of quark Reggeization \cite{FS} in QCD. Till now, this
theory remains less developed than the Reggeized gluon
theory, although a noticeable progress was achieved in
the last years,  in particular, the multi--particle
Reggeon vertices required in the next--to--leading
approximation (NLA) were found \cite{LV}, and the
next--to--leading order (NLO) corrections to the vertices
appearing in the the leading logarithmic approximation
(LLA) were calculated \cite{FF01,KLPV}. All these
calculations were performed assuming the quark
Reggeization hypothesis. However, this hypothesis was not
proved even in the LLA, where merely its
self--consistency was shown, in all orders of $\alpha_s$,
but only in a particular case of elastic quark--gluon
scattering~\cite{FS}. Recently,  the hypothesis was tested
at the NLO in order $\alpha_s^2$ in~\cite{BD-DFG}, where
its compatibility with high-energy behaviour of the
two--loop quark--gluon scattering amplitude was shown and
the NLO correction to the quark trajectory was found in
the limit of the space--time dimension $D\rightarrow4$.
Then, by the explicit two--loop calculations with the help
of $s$--channel unitarity \cite{Bogdan_F_1} the
hypothesis was checked and corresponding correction to
the quark trajectory was found at arbitrary  $D$.

In this paper we suggest a complete proof of the quark
Reggeization hypothesis in the LLA for any quark--gluon
inelastic process in all orders of $\alpha_s$. The proof
is based on the relations required by compatibility of
the multi--Regge  form of QCD amplitudes with the
$s$--channel unitarity (bootstrap relations). We derive
these relations and show that their fulfilment guaranties
the multi--Regge  form. Fulfilment of bootstrap relations
is secured by several conditions (bootstrap conditions)
on the Reggeon vertices and trajectories.  We explicitly
show that these conditions are satisfied by the known
expressions for the vertices and trajectories. The method
of the proof is similar to one used for proving of the
gluon Reggeization in the NLA \cite{Balitsky_LF_1}, but
instead of passing to partial waves we apply recently
introduced operator formalism \cite{FFKP} extended to
consideration of inelastic amplitudes and quark
exchanges.

The paper is organized as follows. In the next Section
necessary denotations, kinematics definition as well as
the form of the multi--Regge inelastic amplitudes are
introduced and explicit expressions for
particle--particle--Reggeon, Reggeon--Reggeon--particle
vertices and quark and gluon trajectories are given. The
bootstrap relations are derived in Section 3. Section 4
is devoted to calculation of the s--channel
discontinuities of the amplitudes.  The bootstrap
conditions for Reggeon vertices and trajectory are
derived in Section 5. In the subsequent Section 6 these
bootstrap conditions are verified. Section 7 concludes
the paper.

\section{The multi--Regge form of QCD amplitudes}

\begin{figure}[h]
\centering
\includegraphics{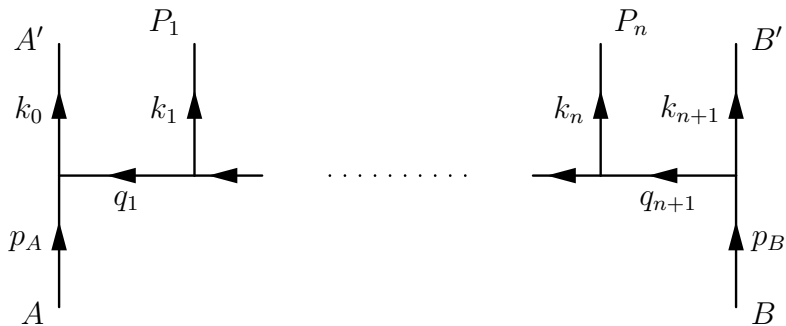}
\caption{Schematic representation of the process $A+B\rightarrow A'+P_1+\cdots +P_n+B'$.}
\label{fig1}
\end{figure}

The only kinematics which is essential in the LLA is the
multi--Regge kinematics (MRK) which means that all
particles participating in a high--energy process  are
well separated in the rapidity space and have limited
transverse momenta.

Let us consider the process $A+B\rightarrow
A'+P_1+....+P_n+B'$ in the MRK. We will use light-cone
momenta $n_1$ and $n_2$,
$\;\;n_1^2=n_2^2=0,\;\;(n_1n_2)=1$,  and denote
$(pn_2)\equiv p^{+},\;\; (pn_1)\equiv
 p^{-}$, so that $pq=p^+q^-+p^-q^++p_\bot q_\bot$, where the sign
 $\bot$ means transverse to the $(n_1, n_2)$
plane components.   We assume that initial momenta $p_A$
and $p_{B}$ (see Fig. \ref{fig1} for denotations) have predominant
components along $n_1$ and $n_2$ respectively.  For
generality we do not assume that transverse  components
$p_{A\bot}$ and $p_{B\bot}$ are zero, but
$|p^2_{A\bot}|\sim |p^2_{B\bot}|\sim p^2_{A} \sim p^2_{B}
\ll p^+_Ap^-_B\;$ and remain limited (do not grow) at
$p^+_Ap^-_B\rightarrow \infty$. For the final particle
momenta $k_i, \;\; i=0, ....,n+1$,  we assume the MRK
conditions:
\begin{align}
&k^-_0\ll k^-_1\ll\ldots\ll k^-_n\ll k^-_{n+1}\,,\quad\nn\\
&k^+_{n+1}\ll k^+_n\ll\ldots\ll k^+_1\ll k^+_0\,,
\label{alph-beta}
\end{align}
and $k_{i\bot}$ are limited. It ensures that the squared
invariant masses $s_{ij}=(k_i+k_j)^2$ are large compared
with the squared transverse momenta; at $i<j$
\begin{equation}
s_{ij}\approx
2k^+_{i}k^-_{j}=\frac{k^+_{i}}{k^+_{j}}(k_j^2-k^2_{j\bot})
=\frac{k^-_{j}}{k^-_{i}}(k_i^2-k^2_{i\bot})\,,\label{s-ij}
\end{equation}
and at $i<l<j$ submit to relations
\begin{equation}
s_{il}s_{lj}\approx
s_{ij}(k_l^2-k^2_{l\bot})\,.\label{s-ilj}
\end{equation}
For the  momentum transfers $q_i$, $i=1,\ldots,n+1$,
\begin{equation}
  q_1=k_0-p_A\,,\;\;  q_{j+1}=q_{j}+k_j\,,\quad(j=1,\ldots,n)\,,\
\end{equation}
we have
\begin{equation}
q_i^2\approx q_\bot^2\,. \label{trans}
\end{equation}
High energy behaviour of amplitudes in the MRK is
determined by exchanges in $q_i$ channels. The largest
$(\sim s_{AB}\equiv (p_A+p_B)^2)$ are  amplitudes with
gluon exchanges in all channels; such exchanges give
factors $ s_i\equiv s_{i-1 i}$ for each of
$q_i$--channel, and product of all these factors gives
$s_{AB}$ due to \eqref{s-ilj}. A quark (antiquark) in a
channel with momentum $q_j$ leads to loss of
$(s_j)^{1/\!2}$.

Our goal is to prove that the amplitude $A_{2\rightarrow
n+2}$ of the process $A+B\rightarrow A'+P_1+....+P_n+B'$
has the multi--Regge form
\begin{equation}\label{inelastic quark}
A^{\mR}_{2\rightarrow n+2} = \bar{\Gamma}^{ \mR_1}_{A' A}
\frac{s_1^{\omega_1}}
{d_1}\gamma^{P_1}_{\mR_1\mR_{2}}\frac{s_{2}^{\omega_2}}
{d_2}.....\gamma^{P_n}_{\mR_i\mR_{n+1}}\frac{s_{n+1}^{\omega_{n+1}}}{
d_{n+1}} \Gamma^{\mR_{n+1}}_{B'B}\,,
\end{equation}
where $\bar{\Gamma}^{\mR}_{A'A}$ and $\Gamma^{\mR}_{B'B}$ are
the particle--particle--Reggeon (PPR) effective vertices,
describing  $P\rightarrow P'$ transitions  due to
interaction with Reggeons $\mR;\;$ for the gluon quantum
numbers in $q_i$ channel $\omega_i=\omega_\mG(q_i)$ is the
gluon  Regge trajectory and $d_i\equiv
d_i(q_i)=q_{i\bot}^2$;   for the quark numbers
$\omega_i=\omega_\mQ(q_i)$ is the quark Regge trajectory
and $d_i\equiv d_i(q_i)=m-\hat{q}_{i\bot};\;$
$\gamma^{P_i}_{\mR_i\mR_{i+1}}$ are the
Reggeon--Reggeon--particle (RRP) effective vertices,
describing  production of particles $P_i$ at Reggeon
transitions $\mR_{i+1}\rightarrow \mR_i$. For definiteness we
do not consider here the antiquark quantum numbers in any
of $q_i$ channels. It determines the order of the
multipliers in \eqref{inelastic quark}. At that, our
consideration does not lose generality, since amplitudes
with quark and antiquark exchanges are related by charge
conjugation.

In order to perform consideration of processes with gluon and
quark exchanges in an unified way we introduced in
\eqref{inelastic quark} denotations slightly different from
usually used. We denote particles and Reggeons by symbols which
accumulate all their quantum numbers. We will use the letter $P$
for  particles and  the letter $\mR$  for Reggeons independently
of their nature, the letters $G$ and $Q$ for ordinary gluons and
quarks and ${\mG}$ and ${\mQ}$ for Reggeized ones. In these
denotations we have for the PPR vertices
\begin{figure}
\centering
\includegraphics{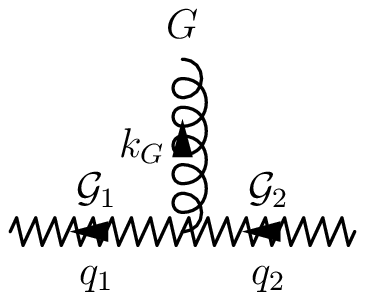}\hfill
\includegraphics{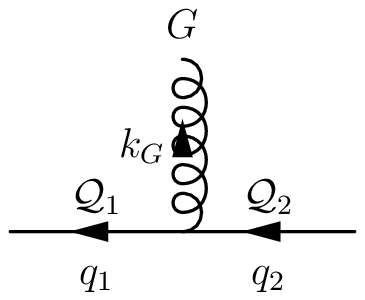}\hfill
\includegraphics{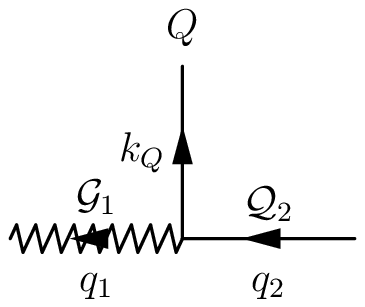}\hfill
\includegraphics{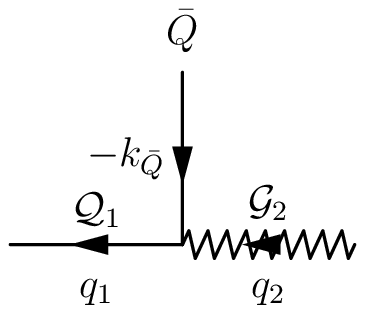}
\\
\hfill a \hfill\hfill b\hfill\hfill c\hfill\hfill d
\caption{The Reggeon--Reggeon--Particle vertices.}
\label{fig2}
\end{figure}
\begin{gather}
\Gamma^{\mG}_{G'G} = -2g\,p_{G}^- {T}^{\mG}_{G'G}(e_{G'_\bot}^\ast
e_{G_\bot})\,,\quad \Gamma^{\mG}_{Q'Q} = g\, \bar{u}_{Q'}t^{\mG}
{\gamma^-}u_Q\,,\quad \Gamma^{\mG}_{\bar Q'\bar Q} = -g\,
\bar{\upsilon}_{\bar Q}t^{\mG}
{\gamma^-}\upsilon_{\bar Q'}\,,\nn\\
\bar{\Gamma}^{\mG}_{G'G} = -2g\,p_{G}^+
{T}^{\mG}_{G'G}(e_{G'_\bot}^\ast e_{G_\bot})\,,\quad
\bar{\Gamma}^{\mG}_{Q'Q} = g\,
\bar{u}_{Q'}t^{\mG}{\gamma^+}u_Q\,,\quad \bar{\Gamma}^{\mG}_{\bar
Q'\bar Q} = -g\, \bar{\upsilon}_{\bar
Q}t^{\mG}{\gamma^+}\upsilon_{\bar Q'}\,, \label{PPRg vertices}
\end{gather}
\begin{gather}
\Gamma^{\mQ}_{G'Q} =- gt^{G'}\hat{e}^\ast_{G'\bot}u_{Q}\,,\quad
\Gamma^{\mQ}_{\bar{Q}'G}
=- gt^{G}\hat{e}_{G\bot}\upsilon_{\bar{Q}'}\,,\nn\\
\bar{\Gamma}^{\mQ}_{G'\bar{Q}} = -
g\bar{\upsilon}_{\bar{Q}}t^{G'}\hat{e}^\ast_{{G'}\bot}\,,\quad
\bar{\Gamma}^{\mQ}_{Q'G} =-
g\bar{u}_{Q'}t^G\hat{e}_{G\bot}\,.\label{PPRq vertices}
\end{gather}
As usually, we do not write colour and spinor quark indices;
${T}^{\mG}$ and  $t^G$ are the color group generators in the
adjoint and fundamental representations. Here and in the following
the physical light--cone gauges
\begin{equation}
\label{n2-gauge} (e_P k_P)=(e_P n_2)=0,\;\;
e_P=e_{P\bot}-\frac{(e_{P\bot}k_P)}{k^+_P}n_2
\end{equation}
and
\begin{equation}
\label{n1-gauge} (e_P k_P)=(e_P n_1)=0,\;\;
e_P=e_{P\bot}-\frac{(e_{P\bot}k_P)}{k^-_P}n_1
\end{equation}
are assumed for polarization vectors $e_P$ of particles
$P$ having momenta $k_P$ with predominant components
along $n_1$ and $n_2$ respectively.

For production of gluon with momentum $k_G=q_2-q_1$ and
polarization vector $e$  in  transition $\mR_2\rightarrow
\mR_1$ of  Reggeons with momenta $q_{2}$ and $q_1$ we have
\cite{Lipatov_1} for the case of Reggeized gluons in both
channels (see Fig. \ref{fig2}a)
\begin{gather}
\gamma^{G}_{{\mG}_1{\mG}_2}=-g{T}^{G}_{{\mG}_1{\mG}_2}\,
{e^\ast}^\mu
C_\mu(q_{2},q_1)\,,\nn\\
C^\mu(q_{2},q_{1})=-(q_{1} + q_2)^\mu_\bot -
{n_1^\mu}\left(k^+_G +\frac{q_{1\bot}^2}{k^-_G}\right) +
{n_{2}^\mu}\left(k^-_G + \frac{q_{2\bot}^2}{k^+_G}\right)\,,
\label{gamma gg}
\end{gather}
and (see Fig. \ref{fig2}b))
\begin{gather}
\gamma^{G}_{{\mQ}_1{\mQ}_2}=-g\, t^{G}\,{e^\ast}^\mu
\mathcal{P}_\mu(q_{2},q_1)\,,\nn\\
\mathcal{P}^\mu(q_{2},q_1) = \gamma^\mu_\bot -
(m-\hat{q}_{{1}\bot})\frac{n^\mu_1}{k^-_G} +
(m-\hat{q}_{2\bot})\frac{n^\mu_{2}}{k^+_G}\, \label{gamma
qq}
\end{gather}
in the case of Reggeized quarks \cite{FS}. It is easy to
check, that these vertices are gauge invariant, since
\begin{equation}\label{c gauge}
    C^\mu(q_{2},q_1)k_{G\mu}=\mathcal{P}^\mu(q_{2},q_1)k_{G\mu}=0\,.
\end{equation}
In the  gauges (\ref{n2-gauge}) and (\ref{n1-gauge})  the
vertices can be presented as
\begin{gather}
 \gamma^{G}_{{\mG}_1{\mG}_2}=2g
 {T}^{G}_{{\mG}_1{\mG}_2}\,
 {e^\ast_\perp}
\left(q_{1\bot}+k_{G\bot}
    \frac{q_{1\bot}^2}{k_{G\bot}^2}\right)\,,\nn\\
\gamma^{G}_{{\mQ}_1{\mQ}_2}=-g t^{G}\,{e^\ast_\perp}
\left(\gamma_\bot-
2(m-\hat{q}_{1\bot})\frac{k_{G\bot}}{k_{G\bot}^2}\right)
\label{gammaG-n2}\,,
\end{gather}
and
\begin{gather}
 \gamma^{G}_{{\mG}_1{\mG}_2}=2g
 {T}^{G}_{{\mG}_1{\mG}_2}\,
 {e^\ast_\perp}
\left(q_{2\bot}-k_{G\bot}
    \frac{q_{2\bot}^2}{k_{G\bot}^2}\right)\,,\nn\\
\gamma^{G}_{{\mQ}_1{\mQ}_2}=-g t^{G}\,{e^\ast_\perp}
\left(\gamma_\bot+
2(m-\hat{q}_{2\bot})\frac{k_{G\bot}}{k_{G\bot}^2}\right)
\label{gammaG-n1}\,
\end{gather}
respectively.

The vertices for quark (antiquark) production were found
in \cite{FS}. For the case of Reggeized gluon in the $q_1
$ channel  (see Fig. \ref{fig2}c) we have
\begin{equation}
\gamma^{Q}_{{\mG}_1{\mQ}_2}=g \, \bar{u}_{Q}
\frac{\hat{q}_{1\bot}}{k^+_Q}t^{\mG_1}\,, \quad\label{gamma q}
\end{equation}
and in the $q_2 $ channel  (see Fig. \ref{fig2}d)
\begin{equation}
\gamma^{\bar Q}_{{\mQ}_1{\mG}_2}=-g t^{\mG_2}\,
\frac{\hat{q}_{2\bot}}{k^-_{\bar{Q}}}\upsilon_{\bar
Q}\,.\quad\label{gamma qbar}
\end{equation}

In terms of integrals in the transverse momentum space
the Reggeon trajectories are presented as
\begin{gather}
    \omega_{\mG}(q)=\frac{N_c}{2}
    \frac{g^2q^2}{(2\pi)^{D-1}}\int
    \frac{\d^{D-2}k_\bot}{k^2_\bot (q-k)^2_\bot}\,,\nn\\
    \omega_{\mQ}(q_i)=C_F
    \frac{g^2}{(2\pi)^{D-1}}(m-\hat{q}_{\bot})\int
    \frac{\d^{D-2}k_\bot}{ (m-\hat{k}_\bot)
    (q-k)^2_\bot}\,,
    \label{trajectory}
\end{gather}
where $N_c=3$ for QCD is number of colours,
$D=4+2\epsilon$ is the space--time dimension taken
different from 4 to regularize infrared divergences.

In the following we will need more general multi--particle
amplitudes $A^{\mR}_{2+n_1\rightarrow 2+n_2}$, but in the same
multi--Regge kinematics. Assuming the same ordering in
longitudinal components, the amplitudes $A^{\mR}_{2+m\rightarrow
2+n-m}$ can be obtained from $A^{\mR}_{2\rightarrow n+2}$ by usual
crossing rules. Note that in \eqref{inelastic quark}  we neglect
imaginary parts of the amplitude since they are subleading.
Therefore the crossing rules for the transition to the amplitudes
do not affect the Regge factors $s_i^{\omega_i}$.

Here it seems sensible to make two remarks. The first one
is that the hypothesis \eqref{inelastic quark} means much
more than it is usually included in the notion
"Reggeization" of elementary particles.  It means not
only  existence of the Reggeons with gluon and quark
quantum numbers and trajectories \eqref{trajectory}, but
also that in the LLA all the MRK amplitudes are
determined only by the Reggeon exchanges, i.e. only
amplitudes with Reggeon  quantum numbers (that means, in
particular, colour octet for pure gluon exchanges and
colour triplet for exchanges with flavour) do survive.
The second remark concerns signature. As compared with
ordinary particles Reggeons possess additional quantum
numbers -- signature, negative for the Reggeized gluon
and positive for the Reggeized quark. Therefore, in order
to affirm that the amplitude is given by the Reggeon
exchanges we need to show that it has corresponding
signatures in all $q_i$--channels.

In order to construct amplitudes with definite signatures
one needs to perform "signaturization". In general the
signaturization is not a simple task. It requires
partial-wave decomposition of amplitudes in
cross-channels with subsequent symmetrization
(anti--symmetrization) in "scattering angles" and
analytical continuation into the $s$--channel. The
procedure is relatively simple only in the case of
elastic scattering of spin-zero particles. At that,
generally speaking, even in this case the amplitudes with
definite signatures can not be expressed in terms of
physical amplitudes related by crossing. Fortunately, at
high energy the signaturization can be easily done not
only for elastic, but in the MRK also for inelastic
amplitudes, for particles with spin as well as for
spin--zero ones. The signaturization (as well as crossing
relations) is naturally formulated for "truncated"
amplitudes, i.e. for amplitudes  with omitted wave
functions (polarization vectors and Dirac spinors). The
crucial points are that in the MRK all energy invariants
$s_{ij}$ are large and that they are determined only by
longitudinal components of momenta ($s_{ij}=2p_i^+p_j^-,
\;\; i<j$). Due to largeness of $s_{ij}$  signaturization
in the $q_l$--channel means symmetrization
(anti--symmetrization) with respect to the substitution
$s_{ij}\leftrightarrow -s_{ij}, \;\; i<l\leq j$. Since
$s_{ij}$ are determined by longitudinal components, it
can be considered as the substitution
$k^{\pm}_i\leftrightarrow -k^{\pm}_i,\;\;i<l,\;\;
p_A^{\pm}\leftrightarrow -p_A^{\pm}$  (or, equivalently,
$k^{\pm}_j\leftrightarrow -k^{\pm}_j,\;\;j\geq l,\;\;
p_B^{\pm}\leftrightarrow -p_B^{\pm}$) in  truncated
amplitudes  without change of transverse components. Note
that such substitution does not violate momentum
conservation due to strong ordering of the longitudinal
components \eqref{alph-beta}. At that, all particles
remain on their mass shell, so that the substitution is
equivalent to transition into the cross-channel.

In order to understand behaviour of the amplitudes
\eqref{inelastic quark} under the signaturization   it is
convenient to take the gluon production vertices in the
physical light--cone gauges with gauge--fixing vectors
$n_2$ or $n_1$ (see \eqref{gammaG-n2},
\eqref{gammaG-n1}). At that, it becomes evident that they
do not depend on longitudinal components of momenta, as
well as the PPR vertices for the Reggeized quark
\eqref{PPRq vertices} after omitting of wave functions.
On the contrary, the quark and antiquark production
vertices \eqref{gamma q} and \eqref{gamma qbar} contain
explicitly longitudinal components, so that they change
their signs at the transition into the cross-channel. The
same is true for the PPR vertices with the Reggeized
gluon \eqref{PPRg vertices}: for the vertices for gluon
scattering because they are proportional to longitudinal
components, and for the vertices for quark and antiquark
scattering because of difference in their signs. After
these remarks, with account of the fact that in the LLA
change of signs of $s_i$ does not affect the Regge
factors,  it is not difficult to see that the amplitudes
\eqref{inelastic quark} are invariant with respect to the
signaturization described above, i.e. they have
corresponding signatures in each of the $q_i$--channels.

\section{Bootstrap relations}
The proof of the form (\ref{inelastic quark}) is based on
use of the {$s$--channel unitarity}, which provides us
with discontinuities $disc_{s_{ij}}$ (i.e. imaginary
parts) of the amplitudes in the $s_{ij}$ channels. We
need to connect the amplitudes themselves (which are real
in the LLA) with these discontinuities. It is not
difficult to do for elastic amplitudes. Unfortunately, it
is quite not so for inelastic amplitudes. Analytical
properties of the production amplitudes are very
complicated even in the MRK \cite{Bartels_1}. But
fortunately, it turns out, that in the LLA these
properties are greatly simplified and allow us to express
partial derivatives $\partial/\partial\ln(s_i)$ of the
amplitudes, considered as a function of $s_i,
\;i=1\ldots n+1,$ and transverse momenta, in terms of the
discontinuities of the signaturized amplitudes. It
permits us to find  all the MRK amplitudes loop by loop
in the perturbation theory, using the Born form of these
amplitudes and the unitarity relations. Note that in the
Born approximation the representation \eqref{inelastic
quark} was proved in \cite{BFKL,FS} with the help
of the $t$--channel unitarity.

For the elastic amplitude  the partial derivative
$\partial/\partial\ln s\;$ can be expressed in terms of
the $s$--channel discontinuity quite easily. For the
signaturized amplitudes radiative corrections depend on
$s$ only in the form $\left(\ln^n(-s)+\ln^n s\right)$
independently of signature. With the LLA accuracy we can
put
\begin{equation}
\frac{1}{-\pi \mathrm{i}}\mathrm{disc}_s \;(\ln^n(-s)+\ln^n
s)=\frac{\partial }{\partial  \ln s}
\left[\ln^n(-s)+\ln^n s\right]~. \label{dis1}
\end{equation}
Therefore we have (the superscript $\mathcal{S}$ means
signaturization)
\begin{equation}
\frac{1}{-\pi \mathrm{i}} \mathrm{disc}_{s}\left[{A}^\mathcal{S}_{2\rightarrow
2}\right]/{A}^\text{Born}_{2\rightarrow
2}=\frac{\partial}{\partial \ln s} \;
\left[{A}^\mathcal{S}_{2\rightarrow
2}/{A}^\text{Born}_{2\rightarrow 2}\right]~.\label{dis2}
\end{equation}
Division by the Born amplitude is performed in order to
differentiate $s$--dependence of radiative corrections
only.

In the case of ${A}_{2\rightarrow 2+n}$ the main
complication is that instead of  $s$ we have
$(n+2)(n+1)/2$ large invariants $s_{ij}=(k_i+k_j)^2$,
which are not independent because of the equalities
\eqref{s-ilj}. Equalities like \eqref{dis2} connecting
discontinuities in each of the channels and corresponding
derivatives of the amplitude do not exist. However there
are  equalities \cite{V.F.2002} connecting definite
combinations of the discontinuities and the derivatives
$\partial/\partial s_i$ :
\begin{equation}\label{bootstrap-0}
\begin{split}
    &\frac{1}{-\pi \mathrm{i}}\left(\sum^{n+1}_{l=k+1}\mathrm{disc}_{s_{kl}}-
    \sum^{k-1}_{l=0}\mathrm{disc}_{s_{lk}}\right)
    A^\mathcal{S}_{2\rightarrow n+2}
    /A^\text{Born}_{2\rightarrow n+2}=\\
    &\left(\frac{\partial}{\partial\ln s_{k+1}}
    -\frac{\partial}{\partial\ln
 s_k}\right)
\left[A^\mathcal{S}_{2\rightarrow
n+2}(s_i)/A^\text{Born}_{2\rightarrow n+2}\right]\,.
\end{split}
\end{equation}
Here in the r.h.s. the amplitude is expressed in terms of
$s_i, \;i=1\ldots n+1,$ and transverse momenta; the index
$k$ takes values from $0$ to $n+1$.

Equalities  \eqref{bootstrap-0} can be easily proved with
use of the Steinmann relations, or, more definitely, of
the statement \cite{Bartels_1} that the amplitude can be
presented as a sum of contributions corresponding to
various sets of $n+1$ nonoverlapping channels
$s_{i_kj_k}\,, \;\;i_k<j_k, \;\; k=1\ldots n+1$; at that
each of the contributions can be written as a
signaturized series in logarithms of energy variables
$s_{i_kj_k}$ with coefficients which are real function of
transverse momenta.  Remind that two channels
$s_{i_1j_1}$ and $s_{i_2j_2}$ are called overlapping if
either $i_1<i_2\leq j_1< j_2$, or $i_2<i_1\leq j_2< j_1$.
What is important:

--- energy variables $s_{i_kj_k}$   are independent, since
the relations \eqref{s-ilj} concern with overlapping
channels; it means, in particular,  that  we need to
consider only leading orders in logarithms of these
variables;

---  we need not to consider the coefficients depending on
transverse momenta  neither calculating  the
discontinuities, nor calculating derivatives over $\ln
s_i$.

Therefore, since scattering amplitudes enter the
relations \eqref{bootstrap-0}  linearly and uniformly, it
is sufficient to prove these relations in the leading
order for the symmetrized products
\begin{equation}
SP=\hat{\mathcal{S}}\prod_{i<j=1}^{n+1}
(-s_{ij})^{\alpha_{ij}}\label{SP}
\end{equation}
instead of $A^\mathcal{S}_{2\rightarrow
2+n}/A^\text{Born}_{2\rightarrow 2+n}$. Here the exponents
$\alpha_{ij}\sim g^2$ are different from zero only for some set of
nonoverlapping channels and are arbitrary in all other respects;
$\hat{\mathcal{S}}$ means symmetrization with respect to
simultaneous change of signs of all $s_{ij}$ with $i<k\leq j$,
performed independently for each $k=1\ldots n+1$. Indeed, due to
above mentioned arbitrariness of $\alpha_{ij}$ fulfilment of
\eqref{bootstrap-0}  for $SP$ guarantees  it for any logarithmic
series.

With $\alpha_{ij}\sim g^2$ calculating  discontinuity of
$SP$ in one of the invariants $s_{ij}$  we can neglect in
the leading order signs of other invariants, so that we
have
\begin{equation}\label{discontinuites}
    \frac{1}{-\pi \mathrm{i}}\left(\sum^{n+1}_{l=k+1}\mathrm{disc}_{s_{kl}}-
    \sum^{k-1}_{l=0}\mathrm{disc}_{s_{lk}}\right)SP=
\left(\sum_{l=k+1}^{n+1}\alpha_{kl}
-\sum_{l=0}^{k-1}\alpha_{lk} \right)SP.
\end{equation}
From other hand,  taking into account that with the LO
accuracy,
\begin{equation}
(s_{ij})^{\alpha_{ij}}=\prod_{l=i+1}^j
s^{\alpha_{ij}}_{l},
\end{equation}
we have
\begin{equation}\label{derivatives}
\left(\frac{\partial}{\partial\ln s_{k+1}}
    -\frac{\partial}{\partial\ln
 s_k}\right)SP=\left(\sum_{i<k+1, j\geq k+1}\alpha_{ij}-\sum_{i<k,
j\geq k}\alpha_{ij}\right)SP
=\left(\sum_{l=k+1}^{n+1}\alpha_{kl}
-\sum_{l=0}^{k-1}\alpha_{lk} \right)SP.
\end{equation}
From \eqref{discontinuites} and \eqref{derivatives} it
follows   that the  equalities  \eqref{bootstrap-0} are
fulfilled.

These equalities allow us to express all partial
derivatives $\partial/\partial\ln(s_k){A}_{2\rightarrow
2+n}$ through the discontinuities. Note that  from $n+2$
equalities \eqref{bootstrap-0} considered as equations
for the derivatives only $n+1$ are linear independent,
that can be easily seen taking sum of the equations over
$k=0\ldots n+1$. Note here that requirement of equality
of mixed derivatives taking in different orders imposes
strong restrictions on the discontinuities. If they are
fulfilled, the amplitude is unambiguously defined by its
value at  $\ln s_i=0$, i.e. in the Born approximation. It
means that the equalities \eqref{bootstrap-0} permit to
find in the LLA all the MRK amplitudes using the Born
approximation for them and the $s$--channel unitarity.
Indeed, at some number $L$ of loops the discontinuities
entering  \eqref{bootstrap-0} can be expressed  with the
help of the $s$--channel unitarity  through the amplitudes
with smaller number of loops. Therefore starting  with
the expression \eqref{inelastic quark} in the Born
approximation (as it was already mentioned, in this
approximation it was proved for arbitrary $n$
\cite{BFKL,FS} with the help of the $t$--channel
unitarity) we can calculate loop--by--loop all radiative
corrections to the Born amplitudes  and examine the
formula \eqref{inelastic quark}.

Instead of such calculations it is sufficient, since the
amplitudes are determined unambiguously, to check that
the Reggeized form \eqref{inelastic quark} satisfies
\eqref{bootstrap-0}. Substituting \eqref{inelastic quark}
into the r.h.s. of \eqref{bootstrap-0} we obtain the
bootstrap relations:
\begin{equation}\label{bootstrap}
\begin{split}
    &\frac{1}{-\pi \mathrm{i}}\left(\sum^{n+1}_{l=k+1}\mathrm{disc}_{s_{kl}}-
    \sum^{k-1}_{l=0}\mathrm{disc}_{s_{lk}}
    \right){A}^\mathcal{S}_{2\rightarrow 2+n}\\
    &={\bar{\Gamma}}^{\mR_1}_{A' A}
   \frac {s_1^{\omega_1}}{ d_1}
    \prod_{i=2}^{k}\left(\gamma^{P_{i-1}}_{\mR_{i-1}\mR_{i}}
\frac{s_{i}^{\omega_i}}{ d_i}\right)
\left[\gamma^{P_{k}}_{\mR_{k}\mR_{k+1}}\omega_{k+1}-\omega_{k}
\gamma^{P_{k}}_{\mR_{k}\mR_{k+1}}\right]
\frac{s_{k+1}^{\omega_{k+1}}}{d_{k+1}}
    \prod_{i=k+2}^{n+1}\left(\gamma^{P_{i-1}}_{\mR_{i-1}\mR_{i}}
\frac{s_{i}^{\omega_i}}{ d_i}\right)
    {\Gamma}^{\mR_{n+1}}_{B'B}\,.
\end{split}
\end{equation}
In the l.h.s. of these equations the discontinuities must
be calculated using the unitarity relations and the
anzats \eqref{inelastic quark}. Since number of the
bootstrap relations is infinite it is quite nontrivial to
satisfy all of them using only several Reggeon vertices
and trajectories. A crucial for the Reggeization
hypothesis fact, which is demonstrated below, is that all
these relations are fulfilled if the Reggeized vertices
and trajectories satisfy several equations called
bootstrap conditions. In the following we derive these
conditions and demonstrate that they are satisfied.

\section{Calculation of the discontinuities}
\begin{figure}[h]
\centering
\includegraphics{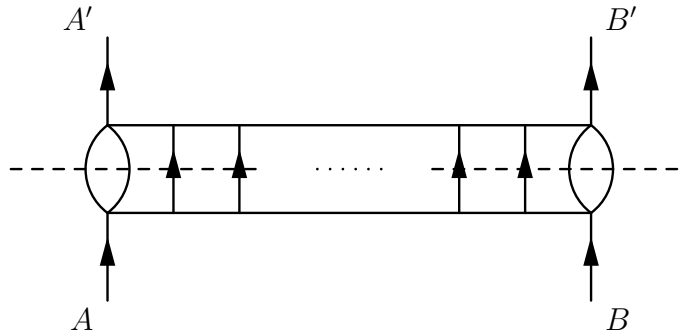}
\caption{Schematic representation of the $s$--channel discontinuity
$\mathrm{disc}_s A^{\mathcal{S}}_{AB\rightarrow A'B'}$.}
\label{fig3}
\end{figure}

Let us start with the elastic amplitude. For the process
${A+B}\rightarrow A'+B'$ the discontinuity is
\begin{equation}\label{elastic discontinuity}
    \mathrm{disc}_{s}A^\mathcal{S}_{{AB}\rightarrow A'B'}=
    \mathrm{i}\sum_{n=0}^\infty   \hat{\mathcal{S}}\int
    A^{\mR}_{{AB}\rightarrow n+2}
    A^{\mR}_{A'B'\rightarrow n+2} \d\rho_{n+2},
\end{equation}
where  $\hat{\mathcal{S}}$ is the signaturization operator, the sum
is taken over discrete quantum states of intermediate particles as
well as over their number, $\d\rho_{n+2}$ is their phase--space
element,  and the hermicity property of the amplitudes
\eqref{inelastic quark} is used. The discontinuity is presented
schematically at Fig. \ref{fig3}, where the  circles on the lines
$AA'$ and $BB'$ mean the signaturization (evidently its execution
for both lines gives the same result as for one of them).

To calculate the discontinuity we need to convolute Reggeon
vertices and to  integrate over momenta of particles in the
intermediate states. All convolutions are known long ago
\cite{BFKL,FS}. The important fact is that they do not
depend on longitudinal momenta. In order to present them, and then
the discontinuities, in a compact way it is convenient to use
operator denotations in the transverse momentum, colour and spin
space. We will use also denotations which accumulate all these
quantum numbers. Thus, $\la{\mG}_i|$ and $|{\mG}_i\ra$ are
"bra"-- and "ket"--vectors for the $t$-- channel states of the
Reggeized gluon with transverse momentum $r_{i \perp}$ and colour
index $c_i$. It is convenient to define the scalar product
\begin{equation}\label{norm G}
\la{\mG}_i|{\mG}_j\ra =r^2_{i \perp}\delta(r_{i
\perp}-r_{j \perp})\delta_{c_ic_j}.
\end{equation}
Analogously, $\la{\mQ}_i|$  and $|{\mQ}_i\ra$ with the
scalar product
\begin{equation}\label{norm Q}
\la{\mQ}_i|{\mQ}_j\ra =(m-\hat{r}_{i
\perp})_{\rho_i\rho_j} \delta(r_{i \perp}-r_{j
\perp})\delta_{\alpha_i\alpha_j}
\end{equation}
denote the $t$-- channel states of the Reggeized quark with
transverse momentum $r_{i \perp}$, colour index $\alpha_i$ and
spinor index $\rho_i$. We will use the letter ${\mR}$  for
denotation of Reggeon states independently of their nature. In the
following we will use the letters ${\mG}_i$ and ${\mQ}_i$ also as
colour indices, instead of $c_i$ and $\alpha_i$. The states with
two Reggeons are built from the above ones. At that it is
convenient to distinguish the states $|{\mR}_i{\mR}_j\ra$
(with corresponding "bra"--vectors $\la{\mR}_i{\mR}_j|$ )and
$|{\mR}_j{\mR}_i\ra$ . We will associate the first of them
with the case when the Reggeon ${\mR}_i$ turns up in the lower
part of Fig. \ref{fig3}, i.e. in the amplitude
$A^{\mR}_{{AB}\rightarrow n+2}$, and the second with the case when
it turns up in the upper part of Fig. \ref{fig3}, i.e. in the
amplitude $A^{\mR}_{n+2 \rightarrow A'B'}$. We define three types
of states
\begin{equation}\label{two-reggeon states}
|{\mG}_i{\mG}_j\ra=|{\mG}_i\ra|{\mG}_j\ra,
\;\;|{\mG}_i{\mQ}_j\ra=|{\mG}_i\ra|{\mQ}_j\ra ,\;\;
|{\mQ}_i{\mG}_j\ra=|{\mQ}_i\ra|{\mG}_j\ra.
\end{equation}
States of different types are  orthogonal one another.
All states create a complete set, i.e.
\[
\la\Psi|\Phi\ra=\int\la\Psi|{\mG}_1{\mG}_2\ra
\frac{\d^{D-2}r_{1\bot}
\d^{D-2}r_{2\bot}}{r^2_{1\bot}r^2_{2\bot}}\la{\mG}_1{\mG}_2|\Phi\ra
+\int\la\Psi|{\mQ}_1{\mG}_2\ra\frac{\d^{D-2}r_{1\bot}
\d^{D-2}r_{2\bot}}{(m-\hat
r_{1\bot})r^2_{2\bot}}\la{\mQ}_1{\mG}_2|\Phi\ra
\]
\begin{equation}\label{completeness}
+\int\la\Psi|{\mG}_1{\mQ}_2\ra\frac{\d^{D-2}r_{1\bot}
\d^{D-2}r_{2\bot}}{(m-\hat
r_{2\bot})r^2_{1\bot}}\la{\mG}_1{\mQ}_2|\Phi\ra,
\end{equation}
where summation over colour and spin indices is assumed.

Interaction of scattering particles with Reggeons is
described by so called impact factors.  We define them as
projections of $t$-channel states $|\bar B' B\ra$ and
$\la A' \bar A|$ on the two-Reggeon states:
\begin{equation}\label{impact b-rr}
\la{\mR}_1{\mR}_2|\bar B' B\ra=\delta(r_{1
\perp}+r_{2\perp}-q_{B
\perp})\frac{1}{2p^-_B}\sum_{P}\left(\Gamma^{{\mR}_2}_{B'P}
\Gamma^{{\mR}_1}_{PB}\;\pm\;{\underline \Gamma}^{{\mR}_2}_{\;\bar
B  P} {\underline \Gamma}^{{\mR}_1}_{\; P\bar B'}\right),
\end{equation}
where the $+$ ($-$) sign stands for a fermion (boson) state in the
$t$--channel, $q_{B}= p_{B}-p_{B'}$, the sum is taken over quantum
numbers of particles $P$ (at that, these particles can be
different in the first and the second terms) and the factor
$1/p^-_B$ is included in the definition for convenience. The
factor $\half$ and the last term in \eqref{impact b-rr} serves for
account of the signaturization; at that, the bar over  particle
symbols means, as usually, antiparticles, while ${\underline
\Gamma}^{{\mR}_2}_{\;\bar B  P}$ and ${\underline
\Gamma}^{{\mR}_1}_{\; P\bar B'}$ are obtained from
${\Gamma}^{{\mR}_2}_{\bar B P}$ and ${\Gamma}^{{\mR}_1}_{ P\bar
B'}$ correspondingly (see \eqref{PPRg vertices}, \eqref{PPRq
vertices}) taking instead of wave functions (polarization vectors
and Dirac spinors) of $\bar B$ and $\bar B'$ the wave functions of
$B$ and $B'$ from the first term.

Quite analogously,
\begin{equation}\label{impact a-rr}
\la{A'\bar A|\mR}_1{\mR}_2\ra=\delta(r_{1
\perp}+r_{2\perp}-q_{A\perp})\frac{1}{2p^+_A}\sum_{P}
\left(\bar{\Gamma}^{{\mR}_2}_{A'P}
\bar{\Gamma}^{{\mR}_1}_{PA}\;\pm\;{\underline
{\bar{\Gamma}}}^{{\mR}_2}_{\;\bar AP} {\underline
{\bar{\Gamma}}}^{{\mR}_1}_{\;P\bar A'}\right),
\end{equation}
where $q_{A}= p_{A'}-p_{A}$.

We  introduce the operator $\hat{\mathcal{K}}_r$  of Reggeon-Reggeon
interaction, related to real particle production.  It is defined
by its matrix elements between the two-Reggeon states, which are
expressed in terms of convolutions of the RRP vertices. The
important remark which must be made here is that, because of the
anticommutativity of the fermion operators the sign of the
amplitude depends on their order in the definition of the state
vectors. We have defined the amplitudes $A^{\mR}$ \eqref{inelastic
quark} without worrying about their signs or fixing this order, as
if the operators were commutative. However in \eqref{elastic
discontinuity} the relative signs of the amplitudes must be taken
into account. In order to do this we must associate a factor -1
with each antiquark in the intermediate state (that can be easily
understood from the Cutkosky  rules). We define:
\begin{equation}\label{kernel operator}
 \la{\mR}_1{\mR}_2|\hat{\mathcal{K}}_r|{\mR}'_1
{\mR}'_2\ra =\delta(q'_{\perp}-q_{\perp})
\frac{1}{2(2\pi)^{D-1}}\sum_{P} \gamma^{P}_{{\mR}_1{\mR}'_1}
    \gamma_{P}^{{\mR}_2{\mR}'_2}\,,
\end{equation}
where $q_{\perp}=r_{1 \perp}+r_{2 \perp}, \;\; q'_{\perp}=r'_{1
\perp}+r'_{2 \perp}$. In this formula we account the above remark
concerning $-1$ for each antiquark in intermediate state by
insertion $-1$ into the definition of the vertex $\gamma_{\bar Q
}^{{\mG}_2{\mQ}'_2}$:
\begin{equation}
\gamma_{\bar Q }^{{\mG}_2{\mQ}'_2}=g \, \bar{\upsilon}_{\bar{Q}}\,
\frac{\hat{p}_{\mG_2\bot}}{k^+_{\bar{Q}}}\,t^{\mG_2}\,,
\end{equation}
\begin{equation}
\gamma_{Q}^{{\mQ}_2{\mG}'_2}=g \,t^{\mG_2'}
\frac{\hat{p}_{\mG'_2\bot}}{k^-_{Q}}\,u_{Q}\,;
\end{equation}
and the vertices $\gamma_{G}^{{\mR}_2{\mR}'_2}$ are obtained from
$\gamma^{G}_{{\mR}_2{\mR}'_2}$ (see \eqref{gammaG-n2}),
\eqref{gammaG-n1}) by the substitution $k_G\rightarrow -k_G$ (in
accordance with momentum conservation) and $e^*_G\rightarrow e_G$.
We introduce also the operator $\om$, so that
\begin{equation}\label{operator Omega}
\om|{\mR}_1{\mR}_2\ra =\left( \omega_{R_1}(r_{1
\perp})+ \omega_{R_2}(r_{2 \perp})\right)|{\mR}_1{\mR}_2\ra\,.
\end{equation}
Denoting momenta of intermediate particles by $k_i$, we
have for the phase space element in \eqref{elastic
discontinuity}
\begin{equation}\label{rho-1}
\d\rho_{n+2}=(2\pi)^{D}\delta(p_A+p_B-\sum_{i=0}^{n+1}
k_i)\prod_{i=0}^{n+1}\frac{\d^{D-1}k_i}{2k_i^0(2\pi)^{D-1}}
=\frac{(2\pi)^D}{p^+_Ap^-_B}\,\delta(p_{A\bot}+p_{B\bot}-
\sum_{i=0}^{n+1} k_{i\bot})\prod_{j=1}^{n}\d y_i
\prod_{i=0}^{n+1}\frac{\d^{D-2}k_{i\bot}}{ 2
(2\pi)^{D-1}}\,,
\end{equation}
where $y_i=\ln {k_i^+}$-- rapidities of the produced
particles, obeying the conditions
\begin{equation}
\ln p_A^+\equiv y_{A}> y_{1}> \ldots > y_{n}> y_B\equiv
-\ln p^-_B\,.\label{ordering}
\end{equation}
Note that we have included the factors $1/p^-_B$ and
$1/p^+_A$ in the definitions of the impact--factors
\eqref{impact b-rr} and \eqref{impact a-rr}, and the
factors $(2(2\pi)^{D-1})^{-1}$ from produced particles
$P_i$ in the definition of the matrix elements of the
kernel \eqref{kernel operator}. Now, taking into account
that with the LLA accuracy $
s_i^{\omega_i}=\e^{\omega_i(y_{i-1}-y_i)}, $ we can
present the discontinuity \eqref{elastic discontinuity}
in the form
\begin{equation}\label{disc-elast}
    \delta(q_{A\bot}-q_{B\bot})\mathrm{disc}_{s}A^{\mR}_{{AB}
\rightarrow A'B'}=\frac{\mathrm{i}}{4(2\pi)^{D-2}}\,
    \la A'\bar A|\hat{G}(Y)|\bar B' B\ra\,,
\end{equation}
where $q_{B\bot}=p_{B\bot}-p_{B'\bot},
\;\;q_{A\bot}=p_{A'\bot}-p_{A\bot}$,$\;Y=y_{A}-y_B$ and
\begin{equation}\label{series for Green operator}
\hat{G}(Y)=  \sum_{n=0}^\infty
\int_{y_B}^{y_{A}}\e^{\om(y_A-y_1)}\d y_1
\hat{\mathcal{K}}_r\int_{y_B}^{y_1}
\e^{\om(y_1-y_2)}\d y_2
\hat{\mathcal{K}}_r....
\int_{y_B}^{y_{n-1}}\e^{\om(y_{n-1}-y_n)}\d y_n
\hat{\mathcal{K}}_r \e^{\om(y_{n}-y_B)}\,.
\end{equation}
It is easy to see that the Green--function operator obeys
the equation
\begin{equation}\label{equation for Green operator}
\frac{d\hat{G}(Y)}{dY}=  \hat{\mathcal{K}}\hat{G}(Y)\,,
\end{equation}
where
\begin{equation}\label{total kernel}
\hat{\mathcal{K}}=\om+\hat{\mathcal{K}}_r\,,
\end{equation}
with initial condition $\hat{G}(0)=1$, so that
\begin{equation}\label{Green exponent}
\hat{G}(Y)=\e^{\hat{\mathcal{K}}Y}=s^{\hat{\mathcal{K}}}\,.
\end{equation}
Eqs. \eqref{disc-elast} and \eqref{Green exponent} give
the  the operator representation of the discontinuities
of elastic amplitudes.

\begin{figure}[h]
\centering
\includegraphics{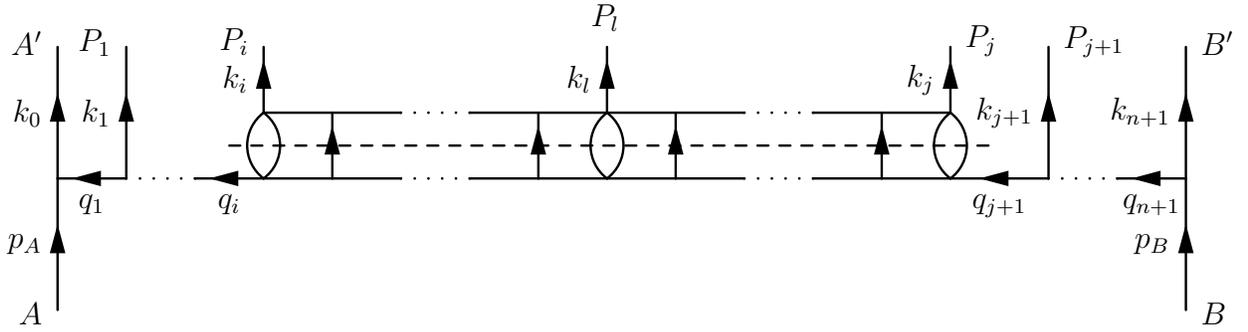}
\caption{Schematic representation of the $s_{ij}$--channel discontinuity
$\mathrm{disc}_{s_{ij}} A^{\mathcal{S}}_{2\rightarrow n+2}$.}
\label{fig4}
\end{figure}
To give analogous representations for discontinuities of
inelastic amplitudes we need to define new operators and
new matrix elements. Let us consider  the discontinuity
schematically presented at Fig. \ref{fig4}, where the circles, as
well as in Fig. \ref{fig3}, mean the signaturization. Analogously
to the impact factors for scattering particles we define
the  impact factors for Reggeon--particle transitions  as
(compare with \eqref{impact b-rr})
\begin{equation}\label{impact rp}
\la{\mR}_1{\mR}_2|\bar P_j \mR_{j+1}\ra=\delta(r_{1
\perp}+r_{2\perp}-q_{j\bot})\frac{1} {2k^-_j}
\sum_{P}\left(\Gamma^{{\mR}_2}_{P_jP} \gamma^{
P}_{{\mR}_1\mR_{j+1}}\;\pm\;{\underline{\Gamma}}^{{\mR}_1}_{\;P\bar
P_j} \gamma_{ P}^{{\mR}_2\mR_{j+1}}\right),
\end{equation}
where $q_{j\bot}=q_{(j+1)\bot}-k_{j\bot}$, the $+$ ($-$)
sign stands for the case of boson (fermion) production;
and
\begin{equation}\label{impact pr}
\la{P_i \mR_{i}|\mR}_1{\mR}_2\ra=\delta(r_{1
\perp}+r_{2\perp}-q_{(i+1)\bot})\frac{1} {2k^+_i}
\sum_{P}\left(\bar{\Gamma}^{{\mR}_2}_{P_iP} \gamma^{
P}_{\mR_{i}{\mR}_1}\;\pm\;{\underline{\bar{\Gamma}}}^{{\mR}_1}_{\;P\bar
P_i} \gamma_{ P}^{\mR_{i}{\mR}_2}\right),
\end{equation}
where $q_{(i+1)\bot}=q_{i\bot}+k_{i\bot}$.

Finally, we introduce the operator $\hat{\mathcal{P}}_l$ for
production of particle $P_l$ with momentum $k_l$ as
having the following matrix elements:
\begin{equation}\label{production operator}
 \la{\mR}_1{\mR}_2|{\hat{\mathcal{P}}}_l|{\mR}'_1
 {\mR}'_2\ra
=\delta(q_{(l+1)\perp}-k_{l\perp}-q_{l\perp}) \left(
\gamma^{P_l}_{\mR_1\mR'_1}\delta_{{\mR}_2{\mR}'_2}
\delta(r_{2\bot}-r'_{2\bot})d_{R_2}+\gamma^{P_l}_{\mR_2\mR'_2}
\delta_{{\mR}_1{\mR}'_1}
\delta(r_{1\bot}-r'_{1\bot})d_{\mR_1}\right)\,,
\end{equation}
where $q_{l\perp}=r_{1 \perp}+r_{2 \perp},\;\;
q_{(l+1)\perp}=r'_{1 \perp}+r'_{2 \perp}$.

Now we are ready to give the operator representation for
discontinuities of the signaturized inelastic amplitudes in the
$s_{ij}$--channels. If $0<i<j<n+1$ (see Fig. \ref{fig4}) then the
value of
$-4\mathrm{i}(2\pi)^{D-2}\delta(q_{i\bot}-q_{(j+1)\bot}-\sum_{l=i}^{j}
k_{l\bot})\;\mathrm{disc}_{s_{ij}}A^\mathcal{S}_{2\rightarrow n+2}$ can be
obtained from the r.h.s of \eqref{inelastic quark} by the
replacement of
\begin{equation}
\gamma^{P_i}_{\mR_i\mR_{i+1}}\left( \prod_{l=i+1}^j
 \frac{s_l^{\omega_l}}{d_l}\gamma^{P_l}_{\mR_l\mR_{l+1}}
 \right)\;\longrightarrow\;
\la P_i,  \mR_i|\left( \prod_{l=i+1}^{j-1}
 s_l^{\hat{\mathcal{K}}}\hat{\mathcal{P}}_l \right)
 s_{j}^{\hat{\mathcal{K}}} |\bar P_j, \mR_{j+1}\ra.
 \label{substitution}
\end{equation}
Eq. (\ref{substitution}) remains valid for $i=0$ with the
substitutions $\gamma^{P_0}_{\mR_0\mR_{1}}\rightarrow
\Gamma^{R_1}_{A^{\prime} A}$ and $\la P_0, \bar{
\mR}_0|\rightarrow \la A^{\prime}, \bar A |$,  as well
as for  $j=n+1$, with the substitutions
$\gamma^{P_{n+1}}_{\mR_{n+1}\mR_{n+2}}\rightarrow
\Gamma^{R_{n+1}}_{B^{\prime} B}$ and $ |\bar P_{n+1},
\mR_{n+2}\ra\rightarrow |\bar B^{\prime}, B \ra$.
The matrix elements  are calculated using the full set of
two-Reggeon states \eqref{two-reggeon states}, the
completeness condition \eqref{completeness} and the
definitions of the operators ${\hat{\mathcal{K}}}$
\eqref{total kernel} and $\hat{\mathcal{P}}_l$
\eqref{production operator} and the matrix elements
\eqref{impact b-rr}, \eqref{impact a-rr}, \eqref{kernel
operator}, \eqref{operator Omega}, \eqref{impact rp}
 and \eqref{impact pr}. After derivation of the
discontinuity \eqref{elastic discontinuity} the
substitution (\ref{substitution}) is practically evident.
In the Born approximation it follows directly from the
above mentioned definitions. The factors $s_i^{\hat{\mathcal{K}}}$ appear from sum of contributions of any number of
intermediate particles with rapidities between $y_{i-1}$
and $y_i$ exactly in the same way as the factor $s^{\hat{\mathcal{K}}}$
 in \eqref{Green exponent}.

It completes the calculation of the discontinuities.

\section{Bootstrap conditions for the Reggeon vertices}

For elastic amplitudes the  bootstrap relation
\eqref{bootstrap} and the representation of the
discontinuity \eqref{elastic discontinuity} give
\begin{equation}\label{disc-elast-2}
    \la{A'\bar A}|\;s^{\hat{\mathcal{K}}}\;|{\bar B'B}\ra=
    -\delta(q_{A\bot}-q_{B\bot})2(2\pi)^{D-1}\bar{\Gamma}^\mR
_{A'A}\omega_\mR
    \frac{s^{\omega_\mR}}{d_\mR}\Gamma^\mR_{B'B}\,,
\end{equation}
where ${\mR}$ takes values ${\mG}$ and ${\mQ}$. This equation
is satisfied if the Reggeon vertices obey the conditions:
\begin{equation}\label{bootstrap vertex}
   |{\bar B'B}\ra=g|{\mR}_{\omega}(q_{B\bot})\ra\Gamma^\mR_{B'B}
    , \;\;\; \la{A'\bar A}|=g\bar{\Gamma}^{\mR}
_{A'A}\la {\mR}_{\omega}(q_{A\bot})|\,,
\end{equation}
where $|\mR_{\omega}(q_{\bot})\ra$ are universal
(process independent) eigenstates  of the kernel
${\hat{\mathcal{K}}}$ with the eigenvalues ${\omega_\mR(q)}$
\begin{equation}\label{bootstrap kernel}
\hat{\mathcal{K}}|{
\mR}_{\omega}(q_\bot)\ra=\omega_\mR(q_\bot)|{
\mR}_{\omega}(q_\bot)\ra\,,\;\; \la{
\mR}_{\omega}(q_\bot)|\hat{\mathcal{K}}=\la{
\mR}_{\omega}(q_\bot)|\,\omega_\mR(q_\bot)\,,
\end{equation}
and with scalar product
\begin{equation}
\la{ \mR}'_{\omega}(q'_\bot)|{
\mR}_{\omega}(q_\bot)\ra
=-\delta_{\mR'\mR}\delta(q'_\bot-q_\bot)C_\mR
\int\frac{\d^{D-2}r_\bot}{d_\mR(r_\bot)(q-r)_\bot^2}\,,
\label{bootstrap norm}
\end{equation}
where $C_\mG=C_A=N_c$,  $\;C_\mQ=2C_F=(N_c^2-1)/N_c$. Note
that the conditions for "ket"-- and "bra"--vectors in
\eqref{bootstrap vertex} and \eqref{bootstrap kernel} are
not independent, because these vectors are related with
each other by the change of $+$ and $-$ momenta
components.

It occurs that  an infinite number of bootstrap relations
for  inelastic amplitudes requires besides
\eqref{bootstrap vertex}--\eqref{bootstrap norm} only one
additional condition. This condition can be obtained from
the bootstrap relation for amplitudes of the process
$A+B\rightarrow A'+P+B'$. Taking in \eqref{bootstrap}
$n=1$  and $k=0$ and writing corresponding
discontinuities according to \eqref{substitution}, we
have
\[
    \la{A'\bar A}|\;s_1^{\hat{\mathcal{K}}}\left(\hat{\mathcal{P}}_{1}
\;s_2^{\hat{\mathcal{K}}}\;|\bar B'B\ra \; +\;|\bar P\mR_2\ra
    \frac{s_2^{\omega_2}}{d_2}\Gamma^{\mR_2}_{B'B}\right)
\]
\begin{equation}\label{disc-inelast}
=-\delta(q_{A\bot}+k_{1\perp}+k_{2\perp}-q_{B\bot})2(2\pi)^{D-1}\bar{\Gamma}^{\mR_1}
_{A'A}\omega_{1}
    \;\frac{s_1^{\omega_1}}{d_1}\;
    \gamma^P_{\mR_1\mR_2}
    \;\frac{s_2^{\omega_2}}{d_2}\;\Gamma^{\mR_2}_{B'B}\,.
\end{equation}
This equality will be satisfied if together with
\eqref{bootstrap vertex}--\eqref{bootstrap norm} the
condition
\begin{equation}\label{bootstrap production ket}
\hat{\mathcal{P}}_{i}\,|\mR_{\omega}(q_{(i+1)\bot})\ra\: g\:
d_{i+1}(q_{(i+1)\bot})+|\bar P_i \mR_{i+1}\ra=
|\mR_{\omega}(q_{i\bot})\ra\,g\:\gamma^{P_i}_{\mR_{i}\mR_{i+1}}
\,,
\end{equation}
where $q_{i\bot}=q_{(i+1)\bot}-k_{i\bot}$,  will be
fulfilled. For the "bra"--vectors this condition is
written as
\begin{equation}\label{bootstrap production bra}
g\: d_{i}(q_{i\bot})\la
\mR_{\omega}(q_{i\bot})|\hat{\mathcal{P}}_{i}\, +\la P_i
 \mR_{i}|= g\:\gamma^{P_i}_{\mR_{i}\mR_{i+1}}\la
\mR_{\omega}(q_{(i+1)\bot})| \,,
\end{equation}
where $q_{(i+1)\bot}=q_{i\bot}+k_{i\bot}$. Let us prove
that the equalities \eqref{bootstrap
vertex}--\eqref{bootstrap norm} and \eqref{bootstrap
production ket}, \eqref{bootstrap production bra}  secure
fulfilment of all infinite set of the bootstrap relations
\eqref{bootstrap}. Consider the terms with $l=n$ and
$l=n+1$ in \eqref{bootstrap}. Corresponding
discontinuities are determined by \eqref{substitution}.
Using \eqref{bootstrap vertex} and \eqref{bootstrap
kernel}  for the $s_{k\,n+1}$--channel discontinuity we
obtain that the sum of the discontinuities in the
channels $s_{kn}$ and $s_{k\,n+1}$ contains
\begin{equation}\label{bootstrap proof 1}
g\:\hat{\mathcal{P}}_{n}\,|\mR_{\omega}(q_{(n+1)\bot})\ra\:
+|\bar P_n \!\mR_{n+1}\ra\frac{1}{d_{n+1}}=
|R_{\omega}(q_{n\bot})\ra\,g\:
\gamma^{P_n}_{\mR_{n}\mR_{n+1}}\frac{1}{d_{n+1}}\,.
\end{equation}
Here the equation \eqref{bootstrap production ket} was
used. Now the procedure can be repeated: we can apply to
this sum Eqs. \eqref{bootstrap vertex} and
\eqref{bootstrap kernel}, and to the sum of the obtained
result with the $s_{k\,n-1}$--channel discontinuity Eq.
\eqref{bootstrap production ket}. Thus all sum over $l$
from $k+1$ to $n+1$ is reduced to one term. Quite
analogous procedure (with use of the bootstrap conditions
for "bra"--vectors) can be applied to the sum over $l$
from $0$ to $k-1$. As a result we have that the left part
of \eqref{bootstrap} with the coefficient $
-2(2\pi)^{D-1}\delta(q_{(k+1)\bot}-q_{k\bot}-k_{k\bot})$,
where $q_{(k+1)\bot}=p_{B\bot}-p_{B'\bot}
-\sum_{l=k+1}^{n} k_{l\bot}, \;\;
q_{k\bot}=p_{A'\bot}-p_{A\bot} +\sum_{l=1}^{k-1}
k_{l\bot}$, can be  obtained from the r.h.s. of
\eqref{inelastic quark} by the replacement
\begin{equation}
\gamma^{P_k}_{\mR_k\mR_{k+1}}\longrightarrow\; \la P_k
\mR_k|\mR_\omega (q_{(k+1)\bot})\ra g d_{k+1}-gd_k\la
\mR_\omega (q_{k\bot})|\bar P_k \mR_{k+1}\ra\,.
\end{equation}
Taking difference of \eqref{bootstrap production ket}
multiplied by $g\,d_i\la \mR_\omega (q_{i\bot})|$ and
\eqref{bootstrap production bra} multiplied by $|
\mR_\omega (q_{(i+1)\bot})\ra g\,d_{i+1}$ and using  the
normalization \eqref{bootstrap norm} we obtain
\[
 \la P_k \mR_k|\mR_\omega
(q_{k+1})\ra\,g\,d_{k+1}-g\, d_k\,\la \mR_\omega
(q_{k})|\bar P_k \mR_{k+1}\ra
=-2(2\pi)^{D-1}\delta(q_{(k+1)\bot}-q_{k\bot}-k_{k\bot})
\]
\begin{equation}
\times
\left(\gamma^{P_k}_{\mR_k\mR_{k+1}}\omega_{\mR_{k+1}}(q_{k+1})-
\omega_{\mR_{k}}(q_{k})\gamma^{P_k}_{\mR_k\mR_{k+1}}\right) .
\end{equation}
It concludes the proof.

Thus, fulfilment of the bootstrap conditions
\eqref{bootstrap vertex}--\eqref{bootstrap kernel} and
\eqref{bootstrap production ket}, \eqref{bootstrap
production bra} secures implementation of all infinite
set of the bootstrap relations \eqref{bootstrap}.

\section{Verification of bootstrap conditions on Reggeon vertices.}

Let us start from the impact factors. As it was already
mentioned, the conditions for "ket"-- and "bra"--vectors
are not independent, so that in the following we
consider only "ket"--vectors. Using the PPR vertices
\eqref{PPRg vertices} and the definition of the vertices
${\underline \Gamma}$ given after \eqref{impact b-rr} it
is easy to obtain
\begin{equation}\label{impact b gg}
\frac{1}{2p^-_G}\sum_{P}\left(\Gamma^{{\mG}_2}_{G'P}
\Gamma^{{\mG}_1}_{PG}\;-\;{\underline \Gamma}^{{\mG}_2}_{\;G P}
{\underline \Gamma}^{{\mG}_1}_{\; P G'}\right)
=-2g^2{T}^{\mG}_{{\mG}_1{\mG}_2}\,p_{G}^-{T}^{\mG}_{G'G}
(e_{G'_\bot}^\ast e_{G_\bot})\,,
\end{equation}
\begin{equation}\label{impact b qq}
\frac{1}{2p^-_Q}\sum_{P}\left(\Gamma^{{\mG}_2}_{Q'P}
\Gamma^{{\mG}_1}_{PQ}\;-\;{\underline \Gamma}^{{\mG}_2}_{\;\bar Q
P} {\underline \Gamma}^{{\mG}_1}_{\; P \bar Q'}\right)
=g^2{T}^{\mG}_{{\mG}_1{\mG}_2}\,\bar{u}_{Q'}t^{\mG}
{\gamma^-}u_Q\,,
\end{equation}
\begin{equation}\label{impact b bar q bar q}
\frac{1}{2p^-_{\bar Q}}\sum_{P}\left(\Gamma^{{\mG}_2}_{\bar Q'P}
\Gamma^{{\mG}_1}_{P\bar Q}\;-\;{\underline \Gamma}^{{\mG}_2}_{\; Q
P} {\underline \Gamma}^{{\mG}_1}_{\; P  Q'}\right)
=-g^2{T}^{\mG}_{{\mG}_1{\mG}_2}\,\bar{\upsilon}_{\bar Q}t^{\mG}
{\gamma^-}\upsilon_{\bar Q'}\,.
\end{equation}
Clearly, in the first of these equations intermediate
particles $P$ are gluons, in the first (second) term of
the second equation they are quarks (antiquarks) and in
the third equation vise versa.  Note that the important
fact of disappearance of all $t$--channel colour states
besides the colour octet one is provided by the
signaturization. All these three equations can be
presented as
\begin{equation}\label{impact b boson}
\frac{1}{2p^-_B}\sum_{P}\left(\Gamma^{{\mG}}_{B'P}
\Gamma^{{\mG}_1}_{PB}\;-\;{\underline \Gamma}^{{\mG}_2}_{\;\bar B
P} {\underline \Gamma}^{{\mG}_1}_{\; P \bar B'}\right) =g
{T}^{\mG}_{{\mG}_1{\mG}_2}\,\Gamma^{{\mG}_2}_{B'B}\,.
\end{equation}
Consequently, according to the definition \eqref{impact
b-rr}, for the case of boson-type $t$--channel states the
bootstrap condition \eqref{bootstrap vertex}  is
fulfilled, and the universal state
$|\mR_{\omega}(q_{\bot})\ra$, which we call in this
case $|\mG_{\omega}(q_{\bot})\ra$, is defined by the
matrix elements
\begin{equation}\label{impact G-gg}
\la{\mG}_1{\mG}_2|\mG_{\omega}(q_{\bot})\ra=\delta(r_{1
\perp}+r_{2\perp}-q_{\bot}){T}^{\mG}_{{\mG}_1{\mG}_2}.
\end{equation}
In the following we will show that this state is the
eigenstate of the kernel ${\hat{\mathcal{K}}}$ with the
eigenvalues ${\omega_\mG(q)}$. Now we turn to the
fermion-type $t$--channel states.

Using the PPR vertices \eqref{PPRq vertices} and
\eqref{PPRg vertices} we obtain
\begin{equation}\label{impact gq bar q g}
\frac{1}{2p^-_G}\sum_{P}\left(\Gamma^{{\mQ}_2}_{\bar Q'P}
\Gamma^{{\mG}_1}_{PG}\;+\;{\underline \Gamma}^{{\mQ}_2}_{\;G P}
{\underline \Gamma}^{{\mG}_1}_{\; P Q'}\right)
=-g^2{t}^{{\mG}_1}t^{G}\hat{e}_{G\bot}\upsilon_{\bar{Q}'}=g
{t}^{{\mG}_1}\Gamma^{{\mQ}_2}_{\bar Q'G} \,.
\end{equation}
Evidently, here in the first term intermediate particles
are gluons and in the second -- quarks. Note that due to
the signaturization only $t$--channel colour triplet does
survive. To obtain \eqref{impact gq bar q g} one needs to
perform commutations of gamma matrices and  to omit
leftmost matrices $\gamma^-$, that can be done due to the
strong ordering \eqref{alph-beta}. The same we will do in
the following with  rightmost $\gamma^+$.

In the same way we obtain
\begin{equation}\label{impact gq gq}
\frac{1}{2p^-_Q}\sum_{P}\left(\Gamma^{{\mQ}_2}_{G'P}
\Gamma^{{\mG}_1}_{PQ}\;+\;{\underline \Gamma}^{{\mQ}_2}_{\;\bar Q
P} {\underline \Gamma}^{{\mG}_1}_{\; P G'}\right)
=-g^2{t}^{{\mG}_1} t^{G'}\hat{e}^\ast_{G'\bot}u_Q = g
{t}^{{\mG}_1}\Gamma^{{\mQ}_1}_{G'Q}\,,
\end{equation}
and
\begin{equation}\label{impact qg bar q g}
\frac{1}{2p^-_G}\sum_{P}\left(\Gamma^{{\mG}_2}_{\bar Q'P}
\Gamma^{{\mQ}_1}_{PG}\;+\;{\underline \Gamma}^{{\mG}_2}_{\;G P}
{\underline \Gamma}^{{\mQ}_1}_{\; P Q'}\right)
=g^2{t}^{{\mG}_2}t^{G}\hat{e}_{G\bot}\upsilon_{\bar{Q}'}=-g
{t}^{{\mG}_2}\Gamma^{{\mQ}_2}_{\bar Q'G} \,.
\end{equation}
\begin{equation}\label{impact qg gq}
\frac{1}{2p^-_Q}\sum_{P}\left(\Gamma^{{\mG}_2}_{G'P}
\Gamma^{{\mQ}_1}_{PQ}\;+\;{\underline \Gamma}^{{\mG}_2}_{\;\bar Q
P} {\underline \Gamma}^{{\mQ}_1}_{\; P  G'}\right)
=g^2{t}^{{\mG}_2} t^{G'}\hat{e}^\ast_{G'\bot}u_Q = -g
{t}^{{\mG}_2}\Gamma^{{\mQ}_1}_{G'Q}\,.
\end{equation}
These equations give
\begin{equation}\label{impact qg b}
\frac{1}{2p^-_B}\sum_{P}\left(\Gamma^{{\mQ}_2}_{B'P}
\Gamma^{{\mG}_1}_{PB}\;+\;{\underline \Gamma}^{{\mQ}_2}_{\;\bar B
P} {\underline \Gamma}^{{\mG}_1}_{\; P B'}\right)  = g
{t}^{{\mG}_1}\Gamma^{{\mQ}_1}_{B'B}\,,
\end{equation}
\begin{equation}\label{impact gq b}
\frac{1}{2p^-_B}\sum_{P}\left(\Gamma^{{\mG}_2}_{B'P}
\Gamma^{{\mQ}_1}_{PB}\;+\;{\underline \Gamma}^{{\mG}_2}_{\;\bar B
P} {\underline \Gamma}^{{\mQ}_1}_{\; P B'}\right)  = -g
{t}^{{\mG}_2}\Gamma^{{\mQ}_1}_{B'B}\,.
\end{equation}
According to the definition \eqref{impact b-rr}, the
bootstrap condition \eqref{bootstrap vertex}  is
fulfilled for the case of fermion-type $t$--channel
states also, with the universal state
$|\mQ_{\omega}(q_{\bot})\ra$, defined by its matrix
elements
\begin{equation}\label{impact Q-gq}
\la{\mG}_1{\mQ}_2|\mQ_{\omega}(q_{\bot})\ra=\delta(r_{1
\perp}+r_{2\perp}-q_{\bot}){t}^{{\mG}_1},
\end{equation}
\begin{equation}\label{impact Q-qg}
\la{\mQ}_1{\mG}_2|\mQ_{\omega}(q_{\bot})\ra=-\delta(r_{1
\perp}+r_{2\perp}-q_{\bot}){t}^{{\mG}_2}.
\end{equation}
Now let us demonstrate that  the  states
$|\mG_{\omega}(q_{\bot})\ra$ and
$|\mQ_{\omega}(q_{\bot})\ra$ are  the eigenstate of the kernel
${\hat{\mathcal{K}}}$ with the eigenvalues ${\omega_\mG(q_{\bot})}$ and
${\omega_\mQ(q_{\bot})}$ correspondingly. First we need to obtain
explicit expressions for matrix elements of the operator
${\hat{\mathcal{K}}_r}$ \eqref{kernel operator}. For matrix elements
between states of two Reggeized gluons  we obtain, using the
vertices \eqref{gamma gg} (actually it is much more convenient to
take them in any of the gauges \eqref{gammaG-n2},
\eqref{gammaG-n1}) and the definition of
$\gamma_{G}^{{\mG}_1{\mG}_2}$ given just after \eqref{kernel
operator}:
\[
  \la{{\mG}_1}{{\mG}_2}|\hat{\mathcal{K}}_r|{{\mG}'_1}
{{\mG}'_2}\ra =\delta(r_{1 \perp}+r_{2 \perp}-r'_{1
\perp}-r'_{2 \perp}) \frac{1}{2(2\pi)^{D-1}}\sum_{G}
\gamma^{G}_{{\mG}_1{\mG}'_1}
    \gamma_{G}^{{\mG}_2{\mG}'_2}
\]
\begin{equation}\label{kernel gg-gg}
=\delta(r_{1 \perp}+r_{2 \perp}-r'_{1 \perp}-r'_{2 \perp})
\mathrm{K}^{GG}_r(r_{1\bot},r_{2\bot};r'_{1\bot},r'_{2\bot}) \frac{2
T^a_{{\mG}_1{\mG}'_1}T^a_{{\mG}'_2{\mG}_2}}{N_c}\,,
\end{equation}

\begin{equation}\label{k gg-gg}
    \mathrm{K}^{GG}_r(r_1,r_2;r'_{1\bot},r'_{2\bot})=\frac{g^2}{(2\pi)^{D-1}}\frac{N_c}{2}
\left((r_{1 \perp}+r_{2 \perp})^2-\frac{r^2_{1\bot}
r^{\prime 2}_{2\bot}+r^2_{2\bot} r^{\prime 2}_{1\bot}}
    {(r_1-r'_1)^2_\bot}\right)\,.
\end{equation}
Now it is easy to see that the state $|{\mG}_\omega
(q_\bot)\ra$ is the eigenstate of $\hat{\mathcal{K}}$ \eqref{total
kernel}. Indeed, since this operator conserves fermion number, it
is sufficient, with account of the completeness condition
\eqref{completeness} and \eqref{impact G-gg}, to show that
\begin{equation}\label{check G omega 1}
  \la{{\mG}_1}{{\mG}_2}|\hat{\mathcal{K}}_r|{{\mG}'_1}
{{\mG}'_2}\ra\la {\mG}'_1 {\mG}'_2|{\mG}_\omega
(q_\bot)\ra =\delta(r_{1 \perp}+r_{2
\perp}-q)\left(\omega_\mG(q_\bot)-\omega_\mG(r_{2\bot})
-\omega_\mG(r_{1\bot})\right)T^\mG_{{\mG}_1{\mG}_2}.
\end{equation}
Using \eqref{completeness}, \eqref{kernel gg-gg} and
\eqref{k gg-gg} it is easy to obtain
\[
\la{{\mG}_1}{{\mG}_2}|\hat{\mathcal{K}}_r|{{\mG}'_1}
{{\mG}'_2}\ra\la {\mG}'_1 {\mG}'_2|{\mG}_\omega
(q_\bot)\ra =\delta(r_{1 \perp}+r_{2 \perp}-q)\frac{N_c}{2}
    \frac{g^2}{(2\pi)^{D-1}}
\]
\begin{equation}\label{check G omega 2}
\times\int
   \frac{\d^{D-2}k_\bot}{ k^2_\bot }\left(\frac{q^2_\bot}
{(q-k)^2_\bot}-\frac{r^2_{1 \perp}} {(r_{1
\perp}-k)^2_\bot}-\frac{r^2_{2\perp}} {(r_{2 \perp}-k)^2_\bot}
\right) T^\mG_{{\mG}_1{\mG}_2}.
\end{equation}
Using the representation \eqref{trajectory} for trajectories in
\eqref{check G omega 1}, we see that it is satisfied, i.e. indeed
$|{\mG}_\omega (q_\bot)\ra$ is the eigenstate of the kernel
with the eigenvalue $\omega_\mG(q_\bot)$.

Turn now to the fermion-type states. The matrix elements
between the states of Reggeized gluon and Reggeized quark
correspond to
\[
  \la{{\mQ}_1}{{\mG}_2}|\hat{\mathcal{K}}_r|{{\mQ}'_1}
{{\mG}'_2}\ra =\la{{\mG}_2}{{\mQ}_1}|\hat{\mathcal{K}}_r|{{\mG}'_2} {{\mQ}'_1}\ra =\delta(r_{1 \perp}+r_{2
\perp}-r'_{1 \perp}-r'_{2 \perp})
\frac{1}{2(2\pi)^{D-1}}\sum_{G}\gamma^{G}_{{\mQ}_1{\mQ}'_1}\gamma_{G}^{{\mG}_1{\mG}'_1}
\]
\begin{equation}\label{kernel qg-qg}
=\delta(r_{1 \perp}+r_{2 \perp}-r'_{1 \perp}-r'_{2 \perp})
\mathrm{K}^{QG}_r(r_{1\bot},r_{2\bot};r'_{1\bot},r'_{2\bot})\frac{2T^a_{{\mG}'_2
{\mG}_2}t^a}{N_c}\,,
\end{equation}
where
\begin{equation}\label{k-G}
    \mathrm{K}^{QG}_{r}(r_{1\bot},r_{2\bot};r'_{1\bot},r'_{2\bot})=
\frac{g^2}{(2\pi)^{D-1}}\frac{N_c}{2}\left(m-\hat{r}_{1\bot}-
\hat{r}_{2\bot}- \frac{(m-\hat{r}_{1\bot})r^{\prime
2}_{2\bot}+(m-\hat{r}^{\prime}_{1\bot})
r^{2}_{2\bot}}{(r_1-r'_1)^2_\bot}\right)\,,
\end{equation}
and
\[
 \la{{\mG}_1}{{\mQ}_2}|\hat{\mathcal{K}}_r|{{\mQ}'_1}
{{\mG}'_2}\ra =\la {\mQ}_2{\mG}_1|\hat{\mathcal{K}}_r|{{\mG}'_2} {{\mQ}'_1}\ra
  =\delta(r_{1 \perp}+r_{2
\perp}-r'_{1 \perp}-r'_{2 \perp}) \frac{1}{2(2\pi)^{D-1}} \sum_{Q}
\gamma^{Q}_{{\mG}_1{\mQ}'_1}\gamma_{Q}^{{\mQ}_2{\mG}'_2}
\]
\begin{equation}
= \delta(r_{1 \perp}+r_{2\perp}-r'_{1 \perp}-r'_{2 \perp})
\mathrm{K}^{GQ}_{r}(r_{1 \perp},r_{2\perp};r'_{1 \perp},r'_{2
\perp})(2N_c){t^{{\mG}'_2}t^{{\mG}_1}}{}\,,\label{kernel gq-qg}
\end{equation}
where
\begin{equation}\label{k-Q}
    \mathrm{K }^{GQ}_{r}(r_{1 \perp},r_{2\perp};r'_{1
\perp},r'_{2
\perp})=\frac{g^2}{(2\pi)^{D-1}}\frac{-1}{2N_c}\left(m-\hat{r}_{1\bot}
- \hat{r}_{2\bot} -
(m-\hat{r}_{2\bot})\frac{1}{m-(\hat{r}'_{1\bot} -
\hat{r}_{1\bot})}
    (m-\hat{r}'_{1\bot})\right)\,.
\end{equation}
In order to prove that the state $|{\mQ}_\omega (q_\bot)\ra$
is the eigenstate of $\hat{\mathcal{K}}$ \eqref{total kernel} we need,
taking into account \eqref{impact Q-gq} and \eqref{impact Q-qg},
to show that
\begin{equation}\label{proof Q}
\la{{\mG}_1}{{\mQ}_2}|\hat{\mathcal{K}}_r|{\mQ}_\omega
(q_\bot)\ra = - \la{{\mQ}_2}{{\mG}_1}|\hat{\mathcal{K}}_r|{\mQ}_\omega (q_\bot)\ra= \delta(r_{1 \perp}+r_{2
\perp}-q)\left(\omega_\mQ(q_\bot)-\omega_\mG(r_{1\bot})
-\omega_\mQ(r_{2\bot})\right)t^{{\mG}_1}.
\end{equation}
Using \eqref{completeness}, \eqref{kernel
qg-qg}--\eqref{k-Q}, and \eqref{impact Q-gq},
\eqref{impact Q-qg},  it is easy to obtain
\[
\la{{\mG}_1}{{\mQ}_2}|\hat{\mathcal{K}}_r|{\mQ}_\omega
(q_\bot)\ra=\la{{\mG}_1}{{\mQ}_2}|\hat{\mathcal{K}}_r|{{\mG}'_1} {{\mQ}'_2}\ra\la {\mG}'_1
{\mQ}'_2|{\mQ}_\omega (q_\bot)\ra
+\la{{\mG}_1}{{\mQ}_2}|\hat{\mathcal{K}}_r|{{\mQ}'_1}
{{\mG}'_2}\ra\la {\mQ}'_1 {\mG}'_2|{\mQ}_\omega
(q_\bot)\ra
\]
\begin{equation}\label{check Q omega}
=\delta(r_{1 \perp}+r_{2 \perp}-q_\perp)
    {g^2}\int
    \frac{\d^{D-2}k_\bot}{2(2\pi)^{D-1}}  \left(\left(
    \frac{(m-\hat q_\bot)}
{(q-k)^2_\bot}-\frac{(m-\hat r_{2\bot})} {(r_2
-k)^2_\bot}\right)\frac{N_c^2-1}{N_c(m-\hat
k_\bot)}-\frac{{N_c}r^2_{1\bot}}{k^2_\bot (r_1 -k)^2_\bot} \right)
t^{{\mG}_1}.
\end{equation}
Together with the representation \eqref{trajectory} it shows that
\eqref{proof Q} is  satisfied, i.e. $|{\mQ}_\omega
(q_\bot)\ra$ is the eigenstate of the kernel with the
eigenvalue $\omega_\mQ(q_\bot)$.

The normalization conditions \eqref{bootstrap norm}
follow  immediately from \eqref{impact G-gg},
\eqref{impact Q-gq}, \eqref{impact Q-qg}.

Thus, we have demonstrated that  the conditions
\eqref{bootstrap vertex}--\eqref{bootstrap norm} are
satisfied. Let us turn now to the last conditions. We
will consider the bootstrap conditions for "ket"--vectors
\eqref{bootstrap production ket} and use the light-cone
gauge \eqref{n1-gauge}. First  we need to find explicit
expressions for the impact factors of Reggeon-particle
transitions and the matrix elements of the production
operator $\hat{\mathcal{P}}_l$  between the vectors
$|\mR_\omega(q_\bot)\ra$ and two-Reggeon states. The
expressions for the impact factors are calculated using
their definition \eqref{impact rp} and  the vertices
\eqref{PPRg vertices}, \eqref{PPRq vertices},
\eqref{gammaG-n1}--\eqref{gamma qbar}.  To find the
matrix elements of the production operator  we use the
definitions of $\hat{\mathcal{P}}_l$ \eqref{production
operator}, the vectors \eqref{impact G-gg}, \eqref{impact
Q-gq}, \eqref{impact Q-qg} and the completeness condition
\eqref{completeness}.

Let us start with gluon production.  In the case of
boson-type  $q_{j+1}$--channel  we obtain for the impact
factor
\[
\la{\mG}_1{\mG}_2|\bar G_j {\mG}_{j+1}\ra=\delta(r_{1
\perp}+r_{2\perp}-q_{j\bot})\frac{1}{2k^-_j}\sum_{G}
\left(\Gamma^{{\mG}_2}_{G_jG} \gamma^{
G}_{{\mG}_1{\mG}_{j+1}}\;+\;{\underline{\Gamma}}^{{\mG}_1}_{\;G
G_j} \gamma_{ G}^{{\mG}_2{\mG}_{j+1}}\right)
\]
\[
=\delta(r_{1
\perp}+r_{2\perp}-q_{j\bot})2g^2e^*_{G_j\perp}\left(-\left(T^{{\mG}_{j+1}}T^{{G}_j}\right)_{{\mG}_1{\mG}_2}\left(q_{(j+1)\bot}-(q_{(j+1)}-r_{1})_\bot
\frac{q^2_{(j+1)\bot}}{(q_{(j+1)}-r_{1})^2_{\bot}} \right)\right.
\]
\begin{equation}\label{impact gg-gg}
\left.+\left(T^{{G}_j}T^{{\mG}_{j+1}}\right)_{{\mG}_1{\mG}_2}\left(q_{(j+1)\bot}-(q_{(j+1)}-r_{2})_\bot
\frac{q^2_{(j+1)\bot}}{(q_{(j+1)}-r_{2})^2_{\bot}} \right)
\right)\,.
\end{equation}
For corresponding martix element of the kernel only
two-gluon intermediate states in the completeness
condition contribute, with the result
\[
 \la{\mG}_1{\mG}_2|\hat{\mG}_j|
\mG_\omega(q_{(j+1)\bot})\ra=\delta(q_{(j+1)\bot}-k_{j
\perp}-q_{j\bot})
2ge^*_{G_j\perp}\left(\left(-T^{{\mG}_{j+1}}T^{{G}_j}\right)_{{\mG}_1{\mG}_2}\left(
\frac{(q_{(j+1)}-r_{1})_\bot}{(q_{(j+1)}-r_{1})^2_{\bot}}-
\frac{k_{j\bot}} {k^2_{j\bot}}\right)\right.
\]
\begin{equation}\label{gg-g-g}
\left.+\left(T^{{G}_j}T^{{\mG}_{j+1}}\right)_{{\mG}_1{\mG}_2}\left(
\frac{(q_{(j+1)}-r_{2})_\bot}{(q_{(j+1)}-r_{2})^2_{\bot}}
-\frac{k_{j\bot}} {k^2_{j\bot}} \right) \right)\,.
\end{equation}
Now, with account of \eqref{impact G-gg},  it is quite
easy to obtain
\begin{equation}
\la{\mG}_1{\mG}_2|\hat{\mathcal{G}}_{i}\,|\mG_{\omega}(q_{(i+1)\bot})\ra\:
g\: d_{i+1}(q_{(i+1)\bot})+\la{\mG}_1{\mG}_2|\bar G_i
 {\mG}_{i+1}\ra= \la{\mG}_1{\mG}_2|\mG_{\omega}(q_{i\bot})\ra\,g\: \gamma^{G_i}_{{\mG}_{i}{\mG}_{i+1}} \,,
\end{equation}
which proves the bootstrap condition \eqref{bootstrap
production ket} for this case.

Another possibility for gluon production is fermion--type
$q_{j+1}$--channel. In this case we have to consider projections
on two different two-Reggeon states: $\la{\mQ}_1{\mG}_2|$ and
$\la{\mG}_1{\mQ}_2|$.  For the first one we obtain
\[
\la{\mQ}_1{\mG}_2|\bar G_j {\mQ}_{j+1}\ra=\delta(r_{1
\perp}+r_{2\perp}-q_{j\bot})\frac{1}{2k^-_j}\left(\sum_{G}
\Gamma^{{\mG}_2}_{G_jG} \gamma^{
G}_{{\mQ}_1{\mQ}_{j+1}}\;+\;\sum_{\bar
Q}{\underline{\Gamma}}^{{\mQ}_1}_{\;\bar Q G_j} \gamma_{ \bar
Q}^{{\mG}_2{\mQ}_{j+1}}\right)
\]
\begin{equation}\label{impact qg-gq}
=\delta(r_{1
\perp}+r_{2\perp}-q_{j\bot})g^2e^*_{G_j\perp}\left([t^{{\mG}_{2}}t^{{G}_j}]\left(\gamma_\bot
+2(m-\hat
q_{(j+1)\bot})\frac{(k_{j}+r_2)_\bot}{(k_{j}+r_2)_\bot^2}\right)
+t^{{G}_j}t^{{\mG}_{2}}\gamma_\bot\frac{1}{m-(\hat k_{j}+\hat
r_1)_\bot}\hat r_{2\bot}\right)\,.
\end{equation}
Calculating corresponding matrix element of the kernel one needs
to take again  only  intermediate states  of one type
($|{\mQ}{\mG}\ra$) in the completeness condition. The result
is
\[
 \la{\mQ}_1{\mG}_2|\hat{\mG}_j|
\mQ_\omega(q_{(j+1)\bot})\ra=\delta(q_{(j+1)\bot}-k_{j
\perp}-q_{j\bot})
ge^*_{G_j\perp}\left(2[t^{{G}_j}t^{{\mG}_{2}}]\left(
\frac{(k_{j}+r_{2})_\bot}{(k_{j}+r_{2})^2_{\bot}}-
\frac{k_{j\bot}} {k^2_{j\bot}}\right)\right.
\]
\begin{equation}\label{qg-g-q}
\left.+t^{{G}_j}t^{{\mG}_{2}}\left(\gamma_\bot +2(m-(\hat
k_{j}+\hat r_1)_\bot)\frac{k_{j\bot}}
{k^2_{j\bot}}\right)\frac{1}{(m-(\hat k_{j}+\hat
r_1)_\bot)}\right)\,,
\end{equation}
so that, with account of \eqref{impact Q-qg},  we  obtain
\begin{equation}
\la{\mQ}_1{\mG}_2|\hat{\mathcal{G}}_{i}\,|\mQ_{\omega}(q_{(i+1)\bot})\ra\:
g\: d_{i+1}(q_{(i+1)\bot})+\la{\mQ}_1{\mG}_2|\bar G_i
 {\mQ}_{i+1}\ra= \la{\mQ}_1{\mG}_2|\mQ_{\omega}(q_{i\bot})\ra\,g\: \gamma^{G_i}_{{\mQ}_{i}{\mQ}_{i+1}} \,,
\end{equation}
so that the bootstrap condition \eqref{bootstrap
production ket} is also fulfilled for this case.

The projection on the state $\la{\mG}_1{\mQ}_2|$ is considered
quite analogously. We obtain
\[
\la{\mG}_1{\mQ}_2|\bar G_j {\mQ}_{j+1}\ra=\delta(r_{1
\perp}+r_{2\perp}-q_{j\bot})\frac{1}{2k^-_j}\left(\sum_{Q}
\Gamma^{{\mQ}_2}_{G_jQ} \gamma^{
Q}_{{\mG}_1{\mQ}_{j+1}}\;+\;\sum_{G}{\underline{\Gamma}}^{{\mG}_1}_{\;G
G_j} \gamma_{G}^{{\mQ}_2{\mQ}_{j+1}}\right)
\]
\begin{equation}\label{impact gq-gq}
=\delta(r_{1
\perp}+r_{2\perp}-q_{j\bot})g^2e^*_{G_j\perp}\left([t^{{G}_j}t^{{\mG}_{1}}]\left(\gamma_\bot
+2(m-\hat
q_{(j+1)\bot})\frac{(k_{j}+r_1)_\bot}{(k_{j}+r_1)_\bot^2}
\right)-t^{{G}_j}t^{{\mG}_{1}}\gamma_\bot\frac{1}{m-(\hat
k_{j}+\hat r_2)_\bot}\hat r_{1\bot}\right)\,,
\end{equation}
and
\[
 \la{\mG}_1{\mQ}_2|\hat{\mG}_j|
\mQ_\omega(q_{(j+1)\bot})\ra=\delta(q_{(j+1)\bot}-k_{j
\perp}-q_{j\bot})
ge^*_{G_j\perp}\left(2[t^{{\mG}_{1}}t^{{G}_j}]\left(
\frac{(k_{j}+r_{1})_\bot}{(k_{j}+r_{1})^2_{\bot}}-
\frac{k_{j\bot}} {k^2_{j\bot}}\right)\right.
\]
\begin{equation}\label{gq-g-q}
\left.-t^{{G}_j}t^{{\mG}_{1}}\left(\gamma_\bot +2(m-(\hat
k_{j}+\hat r_2)_\bot)\frac{k_{j\bot}}
{k^2_{j\bot}}\right)\frac{1}{(m-(\hat k_{j}+\hat
r_2)_\bot)}\right)\,,
\end{equation}
so that, with account of \eqref{impact Q-gq},
\begin{equation}
\la{\mG}_1{\mQ}_2|\hat{\mathcal{G}}_{i}\,|\mQ_{\omega}(q_{(i+1)\bot})\ra\:
g\: d_{i+1}(q_{(i+1)\bot})+\la{\mG}_1{\mQ}_2|\bar G_i
 {\mQ}_{i+1}\ra= \la{\mG}_1{\mQ}_2|\mQ_{\omega}(q_{i\bot})\ra\,g\: \gamma^{G_i}_{{\mQ}_{i}{\mQ}_{i+1}} \,,
\end{equation}
and we see that the bootstrap condition is also
satisfied.

Let us consider now antiquark production. Here again we have to
consider projections on the states $\la{\mQ}_1{\mG}_2|$ and
$\la{\mG}_1{\mQ}_2|$. In the first case we obtain for the
impact factor
\[
\la{\mQ}_1{\mG}_2| Q_j {\mG}_{j+1}\ra=\delta(r_{1
\perp}+r_{2\perp}-q_{j\bot})\frac{1}{2k^-_j}\left(\sum_{\bar Q}
\Gamma^{{\mG}_2}_{\bar Q_j\bar Q} \gamma^{\bar
Q}_{{\mQ}_1G_{j+1}}\;-\;\sum_{
G}{\underline{\Gamma}}^{{\mQ}_1}_{\; G Q_j} \gamma_{
G}^{{\mG}_2{\mG}_{j+1}}\right)
\]
\begin{equation}\label{impact qg-bar qg}
=\delta(r_{1
\perp}+r_{2\perp}-q_{j\bot})\frac{g^2}{k_j^-}\left(t^{{\mG}_{j+1}}t^{{\mG}_{2}}\hat
q_{(j+1)\bot} -[t^{{\mG}_{j+1}}t^{{\mG}_{2}}]\left(\hat
q_{(j+1)\bot} -(\hat k_{j}+\hat
r_1)_\bot\frac{q^2_{(j+1)\bot}}{(k_{j}+r_1)^2_\bot}
\right)\right)\upsilon_{\bar{Q}_j} \,.
\end{equation}
In the matrix element of the kernel now one needs to take
intermediate states  of the  type $|{\mG}{\mG}\ra$. We obtain
\begin{equation}\label{qg-bar q-g}
\la{\mQ}_1{\mG}_2|\hat{\bar{{\mQ}}}_j|
G_\omega(q_{(j+1)\bot})\ra=\delta(q_{(j+1)\bot}-k_{j
\perp}-q_{j\bot}) \frac{g}{k_j^-} [t^{{\mG}_{2}}t^{{\mG}_{j+1}}]
\frac{\hat r_{1\bot}+\hat k_{j\bot}}{(r_{1}+k_{j})_\bot^2}
\upsilon_{\bar{Q}_j}\,,
\end{equation}
so that, with account of \eqref{impact Q-qg}, we see that
the bootstrap condition \eqref{bootstrap production ket}
for this case is also fulfilled,
\begin{equation}
\la{\mQ}_1{\mG}_2|\hat{\bar{{\mQ}}}_j|
\mG_\omega(q_{(j+1)\bot})\ra  g\:
d_{i+1}(q_{(i+1)\bot})+\la{\mQ}_1{\mG}_2| Q_j
{\mG}_{j+1}\ra=
\la{\mQ}_1{\mG}_2|\mQ_{\omega}(q_{i\bot})\ra\,g\: \gamma^{\bar
Q_i}_{{\mQ}_{i}{\mG}_{i+1}} \,,
\end{equation}
which proves the bootstrap condition \eqref{bootstrap
production ket} for this case.

The projection on the $\la{\mG}{\mQ}|$--state is considered
quite similarly. We have
\[
\la{\mG}_1{\mQ}_2| Q_j {\mG}_{j+1}\ra=\delta(r_{1
\perp}+r_{2\perp}-q_{j\bot})\frac{1}{2k^-_j}\left(\sum_{ G}
\Gamma^{{\mQ}_2}_{\bar Q_j G}
\gamma^{G}_{{\mG}_1{\mG}_{j+1}}\;-\;\sum_{
Q}{\underline{\Gamma}}^{{\mG}_1}_{\; Q Q_j} \gamma_{
Q}^{{\mQ}_2{\mG}_{j+1}}\right)
\]
\begin{equation}\label{impact gq-qg}
=\delta(r_{1
\perp}+r_{2\perp}-q_{j\bot})\frac{g^2}{k_j^-}\left([t^{{\mG}_{j+1}}t^{{\mG}_{1}}]\left(\hat
q_{(j+1)\bot} -(\hat k_{j}+\hat
r_2)_\bot\frac{q^2_{(j+1)\bot}}{(k_{j}+r_2)^2_\bot}\right)
-t^{{\mG}_{j+1}}t^{{\mG}_{1}}\hat q_{(j+1)\bot}\right)
\upsilon_{\bar{Q}_j}\,,
\end{equation}
\begin{equation}\label{gq-bar q-g}
\la{\mG}_1{\mQ}_2|\hat{\bar{{\mQ}}}_j|
G_\omega(q_{(j+1)\bot})\ra=\delta(q_{(j+1)\bot}-k_{j
\perp}-q_{j\bot}) \frac{g}{k_j^-}  [t^{{\mG}_{j+1}}t^{{\mG}_{1}}]
\frac{\hat r_{2\bot}+\hat k_{j\bot}}{( r_{2}+ k_{j})_\bot^2}
\upsilon_{\bar{Q}_j}\,,
\end{equation}
and therefore
\begin{equation}
\la{\mG}_1{\mQ}_2|\hat{\bar{{\mQ}}}_j|
\mG_\omega(q_{(j+1)\bot})\ra  g\:
d_{i+1}(q_{(i+1)\bot})+\la{\mG}_1{\mQ}_2| Q_j
{\mG}_{j+1}\ra=
\la{\mG}_1{\mQ}_2|\mQ_{\omega}(q_{i\bot})\ra\,g\: \gamma^{\bar
Q_i}_{{\mQ}_{i}{\mG}_{i+1}} \,.
\end{equation}

Finally, we consider quark production. Here we need to consider
only progection on $\la {\mG}_1{\mG}_2|$--state. It is easy to
obtain
\[
\la{\mG}_1{\mG}_2| \bar Q_j{\mQ}_{j+1}\ra=\delta(r_{1
\perp}+r_{2\perp}-q_{j\bot})\frac{1}{2k^-_j}\left(
\sum_{Q}\Gamma^{{\mG}_2}_{Q_jQ} \gamma^{
Q}_{{\mG}_1\mQ_{j+1}}\;-(-1)\;\sum_{\bar
Q}{\underline{\Gamma}}^{{\mG}_1}_{\;\bar Q \bar Q_j} \gamma_{\bar
Q }^{{\mG}_2\mQ_{j+1}}\right)
\]
\begin{equation}\label{impact qq-gg}
=\delta(r_{1 \perp}+r_{2\perp}-q_{j\bot})\frac{g^2}{k^+_j}\bar
u_{Q_j}\left(t^{{\mG}_{2}}t^{{\mG}_{1}}\hat r_{1\bot}
-t^{{\mG}_{1}}t^{{\mG}_{2}} \hat r_{2\bot}\right) \,,
\end{equation}
where the additional $(-1)$ is introduced to avoid double counting
caused by the fact that the vertices
${\underline{\Gamma}}^{{\mG}_1}_{\;\bar Q \bar Q_j}$ and
$\gamma_{\bar Q }^{{\mG}_2\mQ_{j+1}}$ are both include $-1$ for
antiquark in the intermediate state. Here in the matrix element of
the kernel intermediate states $|{\mQ}{\mG}\ra$ and
$|{\mG}{\mQ}\ra$ contribute. The result is
\[
 \la{\mG}_1{\mG}_2|\hat{\mQ}_j|
\mQ_\omega(q_{(j+1)\bot)}\ra=\delta(q_{(j+1)\bot}-k_{j
\perp}-q_{j\bot}) \frac{g}{k_j^+}\bar u_Q
\]
\begin{equation}\label{gg-q-q}
\times \left(t^{{\mG}_{2}}t^{{\mG}_{1}}\hat r_{2\bot}
\frac{1}{m-(\hat k_{j\bot}+\hat
r_{2\bot})}-t^{{\mG}_{1}}t^{{\mG}_{2}}\hat r_{1\bot}
\frac{1}{m-(\hat k_{j\bot}+\hat r_{1\bot})} \right)\,.
\end{equation}
With account of \eqref{impact G-gg} we  obtain,
\begin{equation}
\la{\mG}_1{\mG}_2|\hat{\mQ}_j|
\mQ_\omega(q_{(j+1)\bot)}\ra g\:
d_{i+1}(q_{(i+1)\bot})+\la{\mG}_1{\mG}_2|\bar{ Q}_j
{\mQ}_{j+1}\ra=
\la{\mG}_1{\mG}_2|\mG_{\omega}(q_{i\bot})\ra\,g\:
\gamma^{Q_i}_{{\mG}_{i}{\mQ}_{i+1}} \,.
\end{equation}
It concludes the proof of the bootstrap relations.

\section{Summary}
The multi--Regge kinematics plays an outstanding role in
high energy physics. It is extremely important since it
gives a dominant contribution to  total cross sections of
particle interactions. The remarkable phenomenon is that
QCD amplitudes in this kinematics have simple
multi--Regge form and are expressed in terms of the gluon
and quark Regge trajectories and a few vertices of
Reggeon interactions.

The multi--Regge form  of amplitudes containing  quark
exchanges  was proposed in \cite{FS} long ago,  but up to
now it was merely tested on its self--consistency for
several particular processes. In this paper we have
presented the proof of the multi--Regge form in the
leading logarithmic approximation for arbitrary
quark--gluon inelastic processes in all orders of
$\alpha_s$. The proof is based on  the bootstrap
relations  required by compatibility of  the multi--Regge
form \eqref{inelastic quark} of inelastic QCD amplitudes
with the $s$--channel unitarity. It consists of three
steps. First, we  derive  an infinite set of the
bootstrap relations \eqref{bootstrap-0} and demonstrate
that  fulfillment of these relations secure the Reggeized
form \eqref{inelastic quark}. Second, we show that  all
these bootstrap relations are  fulfilled if the vertices
and trajectories submit to several bootstrap conditions
\eqref{bootstrap vertex}--\eqref{bootstrap norm},
\eqref{bootstrap production ket} and \eqref{bootstrap
production bra}. This circumstance is extremely
nontrivial since an infinite set of the bootstrap
relations is reduced to several conditions on the Reggeon
trajectories and vertices. And finally, we examine the
bootstrap conditions and prove that all of them are
fulfilled.

Although being simple in principle, necessary
calculations were extremely  cumbersome and tedious if
they performed in the standard approach.  The operator
formalism, recently introduced for consideration of
elastic amplitudes with gluon exchanges and generalized
in this paper for the case of inelastic amplitudes with
arbitrary spin and colour exchanges, is very helpful.

\end{document}